\documentclass{emulateapj}
\pdfoutput=1

\shorttitle{Tidal Origin of the Magellanic Stream}
\shortauthors{Diaz \& Bekki}

\begin{document}

\title{The Tidal Origin of the Magellanic Stream \\ and the Possibility of a Stellar Counterpart}

\author{Jonathan D. Diaz\altaffilmark{1} and Kenji Bekki}
\affil{ICRAR, M468, The University of Western Australia, 35 Stirling Highway, Crawley, Western Australia, 6009, Australia}
\email{jdiaz@ast.cam.ac.uk}

\altaffiltext{1}{Currently at: Institute of Astronomy, University of Cambridge, Madingley Road, Cambridge CB3 0HA, United Kingdom}

\begin{abstract}

We present an N-body model that reproduces the morphology and kinematics of the Magellanic Stream (MS), a vast neutral hydrogen (HI) structure that trails behind the Large and Small Magellanic Clouds (LMC and SMC, respectively) in their orbit about the Milky Way.  After investigating $8\times10^6$ possible orbits consistent with the latest proper motions, we adopt an orbital history in which the LMC and SMC have only recently become a strongly interacting binary pair.  We find that their first close encounter $\sim$2 Gyr ago provides the necessary tidal forces to disrupt the disk of the SMC and thereby create the MS.  The model also reproduces the on-sky bifurcation of the two filaments of the MS, and we suggest that a bound association with the Milky Way is required to reproduce the bifurcation.  Additional HI structures are created during the tidal evolution of the SMC disk, including the Magellanic Bridge, the ``Counter-Bridge," and two branches of leading material.  Insights into the chemical evolution of the LMC are also provided, as a substantial fraction of the material stripped away from the SMC is engulfed by the LMC.  Lastly we compare three different N-body realizations of the stellar component of the SMC, which we model as a pressure-supported spheroid motivated by recent kinematical observations.  We find that an extended spheroid is better able to explain the stellar periphery of the SMC, and the tidal evolution of the spheroid may imply the existence of a stellar stream akin to the gaseous MS.

\end{abstract}


\section{Introduction} \label{sec:int}

Disentangling the interaction history of the Large and Small Magellanic Clouds (LMC and SMC, respectively; ``MCs" collectively) has been facilitated by modeling the formation of the Magellanic Stream (MS), a prominent neutral hydrogen (HI) structure that trails behind the MCs in their orbit about the Milky Way (MW).  Theoretical models have assigned one of two physical mechanisms for creating the MS: tidal stripping (Lin \& Lynden-Bell 1977; Murai \& Fujimoto 1980, MF80; Gardiner, Sawa \& Fujimoto 1994, GSF94; Gardiner \& Noguchi 1996,  GN96; Yoshizawa \& Noguchi 2003; Connors et al. 2006, C06) and ram pressure stripping (Meurer et al. 1985; Heller \& Rohlfs 1994; Moore \& Davis 1994; Mastropietro et al. 2005, M05).  The common feature of these MS formation models is their reliance on multiple strong interactions between the MCs and the Milky Way.  The plausibility of such an interaction history for the MCs has been challenged, however, by the high-precision proper motion measurements of Kallivayalil et al. (2006a,b; K06).  These proper motion estimates imply such large orbital velocities for the MCs that a ``first passage" scenario has been proposed in which the MCs are not bound to the Milky Way but are instead passing by for the first time (Besla et al. 2007).  Because the past orbits of the LMC and SMC have been called into question, the origin of the MS must also be revisited.  { For example, the ``blowout hypothesis" has emerged recently as an alternate scenario for the formation of the MS, which relies on internal mechanisms to expel gas from the MCs (Olano 2004, Nidever et al. 2008).}

Attempts to reconcile the K06 proper motions with the formation of the MS have not yet produced a compelling dynamical model.  { For example, even though the first passage model of Besla et al. (2010, B10) reproduces a tidal tail resembling the MS, the model does not adequately recover the known positions and velocities of the MCs.  In particular, a number of important orbital constraints for the SMC are not satisfied, including on-sky position, line-of-sight velocity, and proper motion (see section~\ref{sec:dx-compare} for a clarifying discussion as well as Besla et al. 2012).  Ruzicka et al. (2010) searched a large parameter space centered on the K06 proper motions, but they were unable to find a test-particle model that could simultaneously explain the on-sky location and kinematics of the MS.  In our previous work, we were able to identify a good MS model, but we had to adopt a massive isothermal halo for the Milky Way in order to combat the large velocities of the MCs (Diaz \& Bekki 2011a,  DB11a).}

{ Nevertheless, one may salvage some insight from the above three studies.  Ruzicka et al. (2010) point out that their most promising models exhibit a common interaction history: two close encounters between the LMC and SMC at $<$2.5 Gyr and 150 Myr ago, each of which triggers an epoch of tidal stripping from the SMC disk.  Similar conclusions are reached in DB11a and B10, in which the MS forms during the first strong binary interaction between the MCs rather than during an interaction with the Milky Way.  Whereas previous tidal models relied on a combination of tides from both the Milky Way and LMC (e.g., GN96, C06), these recent models indicate that the onset of violent LMC tidal interactions could have been sufficient to disrupt the SMC disk and thereby create the MS.}  This new scenario would imply that the binary state of the MCs is a recent phenomenon and that the MS is an artifact from their dynamical coupling.  { Furthermore, observations appear to corroborate this scenario, particularly because the star formation histories of the LMC and SMC exhibit two correlated bursts of star formation at $\sim2$ Gyr ago and $\sim$500 Myr ago (Harris \& Zaritsky 2009).  These two epochs may correspond to two strong tidal interactions suffered mutually by the LMC and SMC, which in turn would create the physical conditions necessary for episodic bursts of cluster formation (Bekki et al. 2004; Piatti et al. 2005; Piatti 2011).}
 
Regarding the tension between the K06 proper motions and MS formation models, a possible compromise has been offered by the recent proper motion measurements of Vieira et al. (2010, V10) and Costa et al. (2009): perhaps the velocities of the MCs are indeed larger than traditionally assumed (e.g., C06, M05) but not as large as what K06 would imply.  The V10 measurements are not nearly as precise as those of K06, but Bekki (2011a) has suggested that the V10 proper motions may nevertheless be more \emph{accurate} than those of K06 due to the larger sample size (3822 stars versus 810 stars, respectively, for the LMC) and the longer baseline (40 years versus 2 years, respectively).  Bekki (2011a) argues that random stellar motions would create unknown systematic errors that could undermine the accuracy of the K06 measurement but would be largely suppressed for V10.  Another unique advantage of V10 is that they measure a precise \emph{relative} proper motion between the LMC and SMC, achieved by tracking the MCs in the same wide field images ($\sim$450 square degrees).  Considering that many properties of the MCs can be understood in terms of their recent activity as a binary pair (star formation histories, Harris \& Zaritsky 2009; bursts of star cluster formation, Bekki et al. 2004, Piatti et al. 2005; formation of MS, DB11a, B10), the V10 constraint on the relative motions of the LMC and SMC are an asset to theoretical models.

In the present work we use orbital models and N-body simulations to address the formation of the MS and other HI structures of the Magellanic system, including the Magellanic Bridge, which extends between the SMC and LMC, and the Leading Arm (LA), which stretches ahead of the MC orbits.  The framework of the present model provides an improvement over previous MS models (e.g., C06, M05, B10) in three important ways.  First, we explore a wide range of orbital histories with velocities constrained by the observational results of V10.  Second, we adopt a multi-component potential for the Milky Way composed of a disk, bulge, and NFW halo, which is a more realistic choice than, for instance, the one-component Milky Way potentials adopted by C06, B10, and DB11a.  And third, we represent the SMC as having both a rotating disk and a non-rotating spheroid, whereas previous MS models have only represented the SMC as a ``pure disk" system.  We are thus able to correctly reproduce the observed kinematics of the SMC, both its rotating HI component (Stanimirovi\'c et al. 2004) and its non-rotating stellar component (Harris \& Zaritsky 2006, HZ06).  Our adopted spheroid model is also consistent with the extended stellar halo of the SMC (e.g., De Propris et al. 2010; Nidever et al. 2011).

In our best model, we find that the SMC disk remains intact until the dynamical coupling of the MCs $\sim$2 Gyr ago, at which point the MS and LA are violently torn away by the strong tidal forces of the LMC.  Additional tidal debris is subsequently engulfed by the LMC, creating a transfer of mass from the SMC that may provide insight into the chemical enrichment history of the LMC.  A second tidal encounter occurring $\sim$250 Myr ago is responsible for pulling the Bridge from the SMC disk as well as a complementary structure that we call the ``Counter-Bridge".  Our model exhibits strong agreement with the HI kinematics and morphology of the MS, particularly its well-known bifurcation into two distinct filaments (Putman et al. 2003a).  For the first time, we are able to provide a structural interpretation of the MS filaments within the context of a dynamical model (however, see Nidever et al. 2008), and we find that a bound association with the Milky Way is required for at least 2 Gyr to explain the bifurcation.  We accordingly suggest that the morphology of the MS provides a strong argument in favor of bound orbits for the MCs, and in particular that a first passage orbit, in which the MCs interact with the Milky Way only recently, may not be able to explain the MS bifurcation.

Using ``disk plus spheroid" N-body models, we describe for the first time the evolution of the SMC spheroid under the same tidal forces that pulled the MS from the SMC disk.  In order to provide a comparative analysis of important physical parameters, three different spheroid models are presented.  We demonstrate that an extended, low-density spheroid is better able to reproduce observations, including the stellar kinematics of the SMC (HZ06), the recent discovery of a break population of red giants (Nidever et al. 2011), and the observation of kinematically and chemically peculiar stars within the LMC (Olsen et al. 2011).  As a consequence of reproducing these observations, we find that the tidal evolution of the extended spheroid necessarily predicts that a coherent stellar stream is stripped away under the same forces that remove the MS and LA from the disk.  The on-sky location of the predicted stellar stream is slightly offset from the observed MS, particularly at its tip, which possibly explains why previous attempts to observe a stellar counterpart to the MS have failed (e.g., Guhathakurta \& Reitzel 1998).

The plan of the paper is as follows: in the next section, we outline our numerical model including a brief discussion of adopted parameters.  We represent the SMC as a multicomponent system (disk, spheroid, and dark matter halo), and we describe the evolution of its disk and its spheroid in separate sections for the sake of clarity.  In section~\ref{sec:disk}, we present our results concerning the tidal evolution of the SMC disk, including a discussion of the MS bifurcation.  In section~\ref{sec:sph} we compare three different spheroid models which enables us to discuss the possibility of a tidal stream of stars.  In section~\ref{sec:dx} we provide a general discussion including a comparison of recent MS formation models.  In section~\ref{sec:con} we summarize and conclude.

\begin{figure*}[t] \centering
\includegraphics[angle=0, scale=0.55, trim=0 100 0 100]{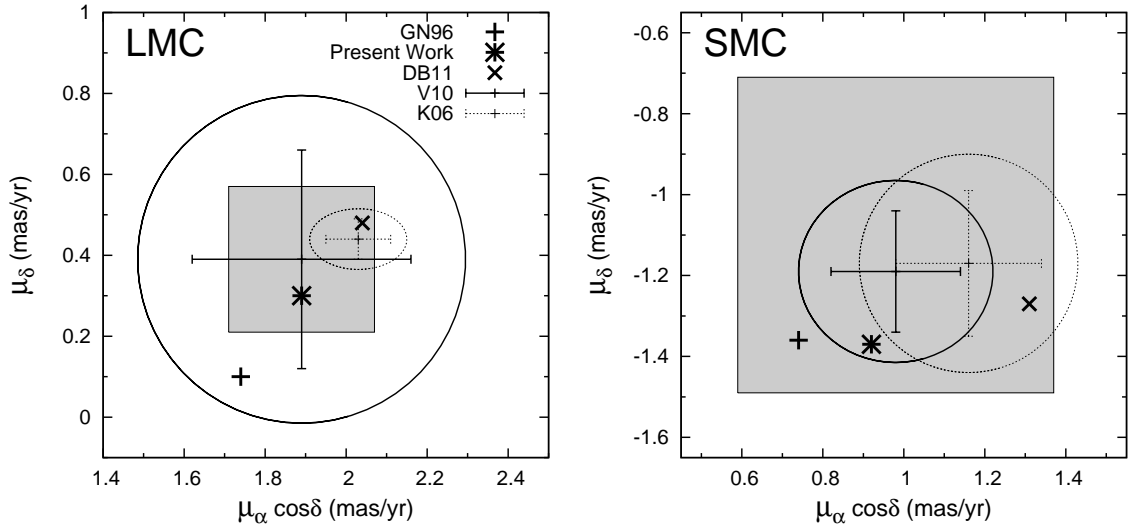}
\caption{The proper motion of the LMC (left) and SMC (right).  Observational ellipses outline the 68.3\% confidence regions and 1$\sigma$ error bars for Vieira et al. (2010; V10, solid) and Kallivayalil et al. (2006a,b; K06, dashed).  The parameter space of possible values explored in the present work are shaded by the grey boxes, and the adopted values are indicated by the star.  The values chosen by Gardiner \& Noguchi (1996; plus sign) and Diaz \& Bekki (2011a; cross) are also shown for comparison.  Models that fail to reproduce the SMC orbit, e.g., Mastropietro et al. 2005 and Besla et al. 2010, are not included in the plot (see text).  The Vieira et. al (2010) proper motion for the SMC is measured \emph{relative} to the LMC, and the corresponding ellipse therefore roams over the large grey box (right panel) as the LMC proper motion is varied.  For the purposes of the figure, we have chosen to fix the ellipse by the LMC proper motion adopted in the present model (star, left panel).
\label{fig:pm}}
 \end{figure*}

\begin{deluxetable*}{lcc}[t]
\footnotesize
\tablecolumns{3}
\tablewidth{0pc}
\tablecaption{Mass and Orbital Parameters
\label{tab:param1}}
\tablehead{
\colhead{  Property } &  \colhead{Adopted Value}  }

\startdata

 & \multicolumn{2}{c}{\bf Milky Way } \\
\\[-2ex]
\hline
\\[-2ex]
 \tablenotemark{$\dagger$}Total Mass ($r<$300 kpc)\tablenotemark{1,2} &  \multicolumn{2}{c}{1.73 $\times$ $10^{12}$ M$_{\odot}$\tablenotemark{$\ast$}}  \\
Disk Mass ($M_{\rm d}$)\tablenotemark{1} &  \multicolumn{2}{c}{5.0 $\times$ $10^{10}$ M$_{\odot}$} \\
Bulge Mass ($M_{\rm b}$)\tablenotemark{1}  &  \multicolumn{2}{c}{0.5 $\times$ $10^{10}$ M$_{\odot}$} \\
NFW Virial Mass ($M_{\rm vir}$)\tablenotemark{2} & \multicolumn{2}{c}{1.30 $\times$ $10^{12}$ M$_{\odot}$\tablenotemark{$\ast$}} \\
NFW Virial Radius ($R_{\rm vir}$) &  \multicolumn{2}{c}{175 kpc\tablenotemark{$\ast$}} \\
Circular Velocity ($V_{ \rm cir })\tablenotemark{3}$ &  \multicolumn{2}{c}{240 km s$^{-1}$\tablenotemark{$\ast$}}  \\
\tablenotemark{$\ddagger$}Distance to Sun ($R_{ \odot }$)\tablenotemark{3,4} &  \multicolumn{2}{c}{8.5 kpc} \\
\tablenotemark{$\ddagger$}Velocity of Sun\tablenotemark{5} &  \multicolumn{2}{c}{(10.0, 5.2 + $V_{ \rm cir }$, 7.2) km s$^{-1}$}  \\

\\[-2ex]
\hline
\\[-2ex]
& {\bf LMC } & {\bf SMC } \\
\\[-2ex]
\hline
\\[-2ex]
 Mass ($M_{\rm mc}$)\tablenotemark{6,7} & $10^{10}$ M$_{\odot}$ & 3 $\times$ $10^{9}$ M$_{\odot}$ \\
Scale radius ($a_{\rm mc}$) & 3 kpc & 2 kpc \\
Proper Motion ($\mu_\alpha \cos\delta $, $\mu_\delta$)\tablenotemark{8} & (1.89, 0.30) mas yr$^{-1}$ &  (0.92, -1.37) mas yr$^{-1}$ \\
Line-of-Sight Velocity\tablenotemark{9,10} & 262 km s$^{-1}$ & 146 km s$^{-1}$ \\
On-Sky Position ($\alpha$, $\delta$)\tablenotemark{9,11} & (81.9$^{\circ}$, -69.9$^{\circ}$)  & (13.2$^{\circ}$, -72.5$^{\circ}$) \\
Distance Modulus\tablenotemark{12,13} & 18.50 & 18.95 \\
\tablenotemark{$\ddagger$}Position Vector ($X$,$Y$,$Z$) & (-0.8, -41.6, -27.0) kpc & (15.3 -36.9 -43.3) kpc \\
\tablenotemark{$\ddagger$}Space Velocities ($U$,$V$,$W$) & (-50.7, -226.1, 229.3) km s$^{-1}$\tablenotemark{$\ast$} & (-4.2, -223.5, 191.0) km s$^{-1}$\tablenotemark{$\ast$} \\

\enddata
\tablenotetext{1}{Binney \& Tremaine (2008);  $^2$Gnedin et al. (2010); $^3$Reid et al. (2009); $^4$Gillessen et al. (2009); $^5$Dehnen \& Binney (1998); $^6$Kim et al. (1998); $^7$Stanimirovi\'c et al. (2004); $^8$Vieira et al. (2010); $^9$van der Marel et al. (2002); $^{10}$Harris \& Zaritsky (2006); $^{11}$Piatek et al. (2008); $^{12}$Freedman et al. (2001); $^{13}$Cioni et al. (2000)}
\tablenotetext{$\dagger$}{Includes bulge mass, disk mass, and NFW halo mass within $r<300$ kpc.}
\tablenotetext{$\ddagger$}{Measured in a galactocentric frame of rest.}
\tablenotetext{$\ast$}{We find that the tidal evolution of the SMC is nearly identical to the present model when the following parameters are substituted for the NFW halo: M$_{\rm vir}$=1.90  $\times$ $10^{12}$ M$_{\odot}$, R$_{\rm vir}$=269 kpc.  In this ``alternate model" (see text), the total mass of the Milky Way within 300 kpc is 2.06 $\times$ $10^{12}$ M$_{\odot}$ and the circular velocity at the solar radius is 220 km s$^{-1}$.  The parameters for the LMC and SMC are same as above, expect for the $V$ component of the space velocities which decrease by 20 km s$^{-1}$ owing to the change in circular velocity.}
\end{deluxetable*}

\section{The Numerical Model}  \label{sec:mod}

The present investigation is three-fold.  First, we search for  the most plausible and realistic orbits of the MCs with respect to the Milky Way by using a backward integration scheme (MF80; GSF94; GN96).  We run $\sim 8\times10^6$ orbital models based on the full range of our parameter space, and we then focus our attention on the subset of models in which the MCs form a recent binary pair.  The recent formation of a strong binary state is assumed to be necessary for the formation of the MS (DB11a).  Second, we run a host of test particle simulations and compare the spatial distributions against that of the observed MS.  This allows us to further narrow the set of promising orbits.  And third, we investigate the structure and kinematics of the simulated MS using medium resolution N-body models.  In this last phase of the investigation, we are able to isolate the best orbital model within our parameter space.  We present our findings as a set of high resolution N-body simulations that explain the observed properties of the MS, LA, Bridge, and stellar structure/kinematics of the SMC in a self-consistent manner.

\subsection{Orbital Models} \label{sec:mod-orb}

Despite the fact that many proper motion studies have been carried out in the last two decades (e.g., see V10 for discussion), an accepted set of measured values has not been converged upon.  Accordingly, we constrain our orbital models by the on-sky positions, distances, and radial velocities of the MCs (see Table~\ref{tab:param1} for adopted values) whereas we explore a large possible range of proper motions.  We center our investigation on the measured values of V10 for the \emph{absolute} proper motion of the LMC, $(\mu_{\alpha} \cos \delta$, $\mu_{\delta})_{\rm LMC}=(1.89\pm0.27$, $0.39\pm0.27)$ mas yr$^{-1}$ and for the more precise \emph{relative} proper motion of the SMC about the LMC  $(\mu_{\alpha} \cos \delta$, $\mu_{\delta})_{\rm SMC}=(\mu_{\alpha} \cos \delta$, $\mu_{\delta})_{\rm LMC}-(0.91\pm0.16$, $1.49\pm0.15)$ mas yr$^{-1}$.  Figure~\ref{fig:pm} gives the investigated range as grey boxes for the LMC (left panel) and for the SMC (right panel).  Notice that as we select different values for the LMC proper motion, the corresponding V10 constraint on the SMC (solid ellipse, right panel) roams across the plane to different values.  The range of proper motions that we consider is sufficiently large to overlap with the 1$\sigma$ errors of the K06 measurement, shown as a dotted ellipse for the LMC (left) and SMC (right).  To avoid clutter in the figure we plot only the observed values of V10 and K06, but other recent measurements deserve to be mentioned such as Piatek et al. (2008) and Costa et al. (2009).

The adopted proper motion values of the LMC and SMC in the present work are indicated by the stars in Figure~\ref{fig:pm}.  We selected these particular values after considering $\sim 8 \times 10^6$ orbital models for the MCs (see below for details).  The adopted values fall between the previously assumed values of GN96 (plus sign) and DB11 (cross).  In this sense, the present work is a compromise between the classic bound scenario for the MCs (i.e., GN96) and the high velocity orbits implied by K06.  There are a host of tidal models for the formation of the Magellanic Stream which use the same proper motions assumed by GN96, including GSF94, Gardiner (1999), Yoshizawa \& Noguchi (2003), and Connors et al. (2006).  The plus sign in Figure~\ref{fig:pm} represents all of these previous works.  Models that reproduce \emph{only} the LMC orbit have been excluded from the plot, including the ram pressure stripping model of Mastropietro et. al (2005), which omits the SMC entirely, and the first passage model of Besla et al. (2010), which does not reproduce the correct SMC position or velocity (see Besla et al. 2012).

We calculate the past orbits of the LMC and SMC with respect to the Milky Way by adopting the backward integration scheme originally devised by MF80.  To carry out these orbit integrations, we must assume model parameters for the following quantities: (1) the shape of the Milky Way's gravitational potential as a function of distance $r$ from the Galactic center, (2) gravitational potential of the MCs, (3) total masses of the MCs, and (4) the form of dynamical friction between the Milky Way's dark matter halo and the LMC (SMC).  For a given set of proper motions, we investigate a range of orbital models based on different parametrizations.

We assume that the Milky Way influences the orbits of the MCs through a \emph{fixed} gravitational potential of three components: a central bulge, a disk, and an extended dark matter halo.  For the halo, which is the dominant component of the MW at the large distances of the MCs, we have adopted an NFW density distribution (Navarro et al. 1996) suggested from CDM simulations:
\begin{equation}
{\rho}_{\rm NFW} (r)=\frac{\rho_{0}}{(cr/R_{\rm vir})(1+cr/R_{\rm vir})^2},
\end{equation}
where $r$ is the spherical radius, $\rho_{0}$ is the characteristic density, $c$ is the concentration parameter, and $R_{\rm vir}$ is the virial radius.  The total mass within $r=R_{\rm vir}$ is called the virial mass and is given by
\begin{equation}
M_{\rm vir}=4\pi \rho_{0} (R_{\rm vir}/c)^3( \ln(1+c) - c/(1+c)).
\end{equation}
For halos with a mass comparable to that of the MW, typical values for the concentration parameter fall in the range $c=10-17$ (Klypin et al. 2002), and we choose $c=12$ in the present study.  We consider $M_{\rm vir}$ and $R_{\rm vir}$ to be free parameters, allowing us to investigate different models for the orbital evolution of the MCs within the halo of the MW.  In particular, we searched for halos within the range $M_{\rm vir}=1-2$ $\times$ $10^{12}$ M$_{\odot}$.

The disk of the Milky Way is represented by a Miyamoto-Nagai (1975) potential
\begin{equation}
{\Phi}_{\rm disk}=-\frac{GM_{\rm d}}{\sqrt{R^2 +{(a+\sqrt{z^2+b^2})}^2}},
\end{equation}
where $M_{\rm d}$ is the total mass of the disk, $a$ and $b$ are scale parameters that control the radial and vertical extent of the disk, respectively, and $R=\sqrt{x^2+y^2}$.

We adopt a spherical Hernquist (1990) model for the potential of the Galactic bulge,
\begin{equation}
{\Phi}_{\rm bulge}=-\frac{GM_{\rm b}}{r+c_{\rm b}},
\end{equation}
where $M_{\rm b}$ and $c_{\rm b}$ are the total mass and the scale length of the bulge, respectively.

\begin{figure} \centering
\includegraphics[trim=55 50 0 370, scale=0.5]{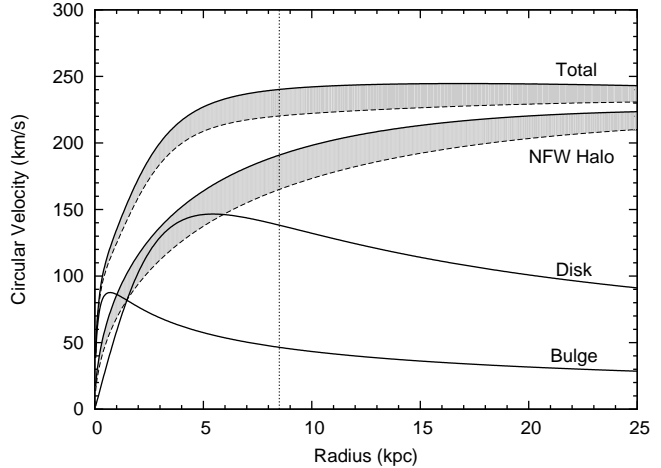}
\caption{The rotation curve of the Milky Way, shown for its individual components (Bulge, Disk, NFW Halo) and the sum of the components (Total).  A range of NFW halos were investigated, represented by the grey shaded regions.  The adopted Milky Way model is given by the upper bound to the shaded regions (solid line; NFW halo $M_{\rm vir}=1.30 \times 10^{12}$~M$_{\odot}$, $R_{\rm vir}=$175 kpc), while an alternate model (see text) is indicated by the lower bound of the shaded regions (dashed line; NFW halo $M_{\rm vir}=1.90 \times 10^{12}$~M$_{\odot}$,  $R_{\rm vir}=$269 kpc).  The vertical dotted line at $R=8.5$ kpc indicates the radius of the Sun, and its intersection with the Total rotation curve accordingly gives the circular velocity $V_{\rm cir}$ value.
\label{fig:rot}}
 \end{figure}

In the present study, the above parameters for the disk and bulge are fixed to the following values (see also Table~\ref{tab:param1}): $M_{\rm d}=5.0 \times$ $10^{10}$ M$_{\odot}$, $a=3.5$ kpc, and $b=0.35$ kpc for the disk; $M_{\rm b}=0.5 \times$ $10^{10}$ M$_{\odot}$ and $c_{\rm b}=0.7$ kpc for the bulge (Binney \& Tremaine 2008).  That is, we did not change these values in our parameter space search for the best tidal model.  Even though the disk and bulge do not influence the orbits of the MCs as strongly as the dark matter halo, the bulge and disk are nevertheless important in determining the circular velocity $V_{\rm cir}$ of the Milky Way (see Fig~\ref{fig:rot}).  We have investigated models having the IAU standard value $V_{\rm cir}=220$ km s$^{-1}$ (Kerr \& Lynden-Bell 1986), and we have also considered a perhaps more realistic value of $V_{\rm cir}=240$ km s$^{-1}$ (Reid et al. 2009; Reid \& Brunthaler 2004; Sirko et al. 2004).  By varying the halo parameters $M_{\rm vir}$ and $R_{\rm vir}$ and considering the added affect of the bulge and disk, we can construct a variety of plausible MW models for our investigation of MC orbits.

The parameter values of our best model are summarized in Table~\ref{tab:param1}.  Our chosen NFW halo has $M_{\rm vir}=130$ $\times$ $10^{10} M_{\odot}$ and $R_{\rm vir}=175$ kpc, which when added to the bulge and disk gives a total mass of 1.73 $\times$ $10^{12}$ M$_{\odot}$ within $r=300$ kpc of the Milky Way.  The circular velocity is $V_{\rm cir}=240$ km s$^{-1}$.  In the following sections we will describe the results of this model, but first we briefly mention an ``alternate model'' for the Milky Way in which the evolution of the MCs is remarkably similar.  In particular, the results of sections~\ref{sec:disk} and~\ref{sec:sph} are largely unchanged when we substitute M$_{\rm vir}$=190 $\times$ $10^{10}$ M$_{\odot}$ and R$_{\rm vir}$=269 kpc for the NFW halo.  In this model, the total mass of the Milky Way within 300 kpc is 2.06 $\times$ $10^{12}$ M$_{\odot}$ and the circular velocity at the solar radius is 220 km s$^{-1}$.  The predictions of this alternate model may be so similar in part because the total mass within the current location of the MCs $r<55$ kpc is identical to that of the adopted model, 6.7 $\times$ $10^{11}$ M$_{\odot}$.  The rotation curves of the adopted and alternate models are given in Fig~\ref{fig:rot}.  { The alternate model highlights the important fact that our chosen parameterization of the Milky Way is not unique and in fact is quite flexible when considering the orbital evolution of the MCs.  Further details of the alternate model are given in the Appendix.}

For the purposes of orbit integration, the LMC and SMC are each assumed to have a Plummer potential of the form
\begin{equation}
 {\Phi_{\rm mc}}(r)=-GM_{\rm mc}/\sqrt{{r}^2+a_{\rm mc}^2},
\end{equation}
where $M_{\rm mc}$ is the total mass of either the LMC or SMC (MC), $r$ is the distance from the center of mass, and $a_{\rm mc}$ is the scale radius of the MC.  In our parameter space search, we fixed the values of the scale radii, but we allowed the masses to vary as free parameters.  Table~\ref{tab:param1} gives the adopted values.

Dynamical friction is assumed to operate separately on the orbits of the LMC and SMC as they pass through the Milky Way's dark matter halo.  We account for this effect by adopting the Chandrasekhar formula (Binney \& Tremaine 2008):
\begin{equation} \label{chandra}
F_{\rm fric, G}= -
\frac {4 \pi {\rm G}^2 M_{\rm mc}^2 \ln(\Lambda) \rho_{\rm dm}  (r) } { v^2} 
[ {\rm erf}(X) - \frac {2X} {\sqrt{\pi}} \exp (-X^2) ] 
\frac { {\bf v} } { v },
\end{equation}
where $M_{\rm mc}$ is  the mass of either the LMC or the SMC (MC), $v$ is the velocity of the MC, and $X=v/(\sqrt{2}\sigma)$, where $\sigma$ is the one-dimensional velocity dispersion of the adopted dark matter halo.  As the MC moves through the halo, we calculate $\sigma$ at a given position using the analytical approximation derived by Zentner \& Bullock (2003).  We adopt a reasonable value of 3.0 for the Coulomb logarithm ${\Lambda}$ (GN96).

We carry out our simulations in a galactocentric frame such that the center of the Milky Way is always set to be ($X$,$Y$,$Z$) = (0,0,0).  The initial position vectors and space velocities of the MCs are calculated from our choice of free parameters (such as proper motions and $V_{\rm cir}$; see Table~\ref{tab:param1} for adopted values) according to the method described in section~9 of van der Marel et al. (2002).  We then integrate the appropriate equations of motion from the present epoch to 5 Gyr in the past using the method devised by MF80.  By varying our free parameters, we are able to investigate a large number $\sim 8 \times 10^6$ of observationally constrained orbital models for the MCs.

A number of studies suggest that the LMC and SMC became dynamically coupled only recently (Harris \& Zaritsky 2009; Bekki et al. 2004), and in our previous work we explored the idea that the first strong interaction during this coupling was responsible for creating the Magellanic Stream (DB11a).  Guided by these results, we restrict our attention to the MC orbital models that satisfy an assumed set of physical conditions for the formation of the Magellanic Stream.  Namely, those assumed conditions are: (i) the mutual separation of the MCs has been less than 30 kpc for the majority of the last 2-3 Gyr, allowing them to dynamically interact, (ii) they strongly interact only twice, at $t=-2.0\pm0.25$ Gyr and $t=-0.4\pm0.25$ Gyr (Harris \& Zaritsky 2009), and (iii) their mutual separation cannot be too small ($> 5$ kpc) at each epoch of strong interaction.  These three requirements enable us to reduce the number of possible orbital models to $\sim 10^5$.  The models within this reduced set are further explored using test particle simulations in the next stage of investigation.

\begin{figure} \centering
\includegraphics[trim=55 50 0 370, scale=0.5]{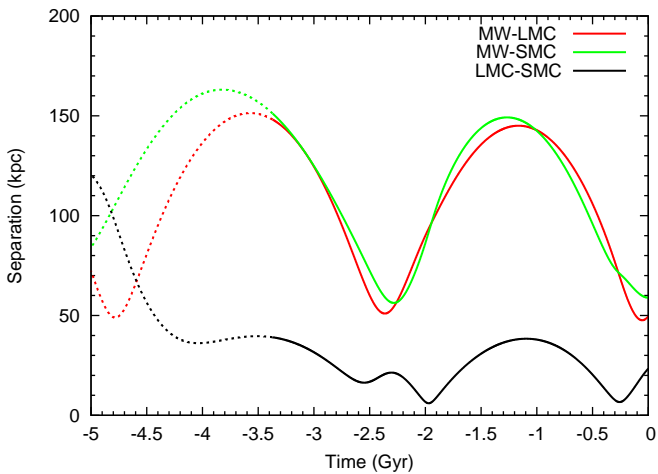}
\caption{Orbital separations between the LMC and MW (red), SMC and MW (green), and LMC and SMC (black).  Solid lines indicate the time window in which the N-body models are evolved, $t\geq-3.37$ Gyr.  Dotted lines indicate the continuation of the orbit integrations to 5 Gyr in the past.
\label{fig:sep}}
 \end{figure}

 \begin{deluxetable*}{cccc}[t]
\footnotesize
\tablecolumns{3}
\tablewidth{0pc}
\tablecaption{SMC N-body Parameters
\label{tab:param2}}
\tablehead{
\colhead{} & \colhead{Halo} & \colhead{Disk} & \colhead{Spheroid}}
\startdata
Model 1 & $N_{\rm dm}$=500,000 & $N_{\rm d}$=400,000 & $N_{\rm sph}$=100,000 \\
& $M_{\rm dm}$=$1.36 \times 10^9$ M$_{\odot}$ & $M_{\rm d}$=$1.36 \times 10^9$ M$_{\odot}$ & $M_{\rm sph}$= $0.27 \times 10^9$ M$_{\odot}$\\
& $R_{\rm dm}$=7.5 kpc & $R_{\rm d}$=5.0 kpc & $R_{\rm sph}$=7.5 kpc \\
& - & ${\theta}_{\rm d}=-45^{\circ}$ ; ${\phi}_{\rm d}=230^{\circ}$ & - \\

\\[-2ex]
\hline
\\[-2ex]
Model 2\tablenotemark{$\ast$} & $R_{\rm dm}$=5.0 kpc & $R_{\rm d}$=5.0 kpc & $R_{\rm sph}$=5.0 kpc \\
\\[-2ex]
\hline
\\[-2ex]
Model 3\tablenotemark{$\ast$} & $R_{\rm dm}$=5.0 kpc & $R_{\rm d}$=5.0 kpc & $R_{\rm sph}$=2.5 kpc \\
\enddata
\tablenotetext{$\ast$}{Models 2 and 3 have the same parameters as Model 1 except for those noted.}
\tablenotetext{1}{For brevity, variables are written without the subscript `SMC' as in the text.}
\end{deluxetable*}

\subsection{Test Particle Simulations} \label{sec:mod-test}

Using the set of orbits derived in the previous step, we run a large number of test particle simulations (e.g., see DB11) to determine which models are best able to reproduce the observed structure of the Magellanic Stream.  Because our focus is mainly on the tidal evolution of the SMC, we represent the SMC by a disk of test particles and leave the LMC as a fixed Plummer potential, as described previously.  The SMC disk is assumed to have an exponential profile with a truncation radius of $R_{\rm d, SMC}$, scale length of $R_{\rm 0, SMC}=0.2R_{\rm d, SMC}$, and a scale height of $Z_{\rm 0, SMC}=0.02R_{\rm d, SMC}$.  Each test particle in the disk is given an initial circular velocity according to the rotation curve derived from the adopted Plummer potential for the SMC.  In order to allow the investigation of a large number of models, we use a coarse resolution of 2000 total particles in the test particle disk.  This number is however more than enough to determine which orbital models provide promising tidal stripping scenarios for the MS.

{ This phase of the study resembles the work of Ruzicka et al. (2010), who also perform a large suite of test particle models in search of a formation scenario for the MS.  However, the presently explored range of proper motions is much larger (Fig~\ref{fig:pm}) than that of Ruzicka et al. (2010), who restrict their attention to the 1$\sigma$ region of K06 and Piatek et al. (2008).  Another key difference is that Ruzicka et al. (2010) use an automated genetic algorithm to determine which models in their parameter space are best.  In contrast, we use an \emph{ad hoc} set of filters to intuitively search through the candidate test particle models.  In particular, we} search for models that can satisfy the following requirements: (i) about 50\% of the stripped SMC particles are located along the observed MS and (ii) more than 30\% yet less than 70\% of the SMC particles are stripped.  { Inspecting the models by eye verified that these criteria could serve as good filters, but they are by no means unique or optimal.  Our intent is to simply show how we arrived at our preferred model.}

Imposing the above criteria enabled us to further narrow the parameter space, as only $\sim$10\% of the test particle models pass these conditions.  Of these models, most fall into one of several ``families" which exhibit similar tidal stripping of the SMC disk.  We take the best candidates from these families and pass them to higher resolution N-body simulations to provide more faithful dynamical models.

\subsection{N-body  Simulations} \label{sec:mod-nbody}

In this final phase of the investigation, we represent the SMC by a system of N-body particles whereas the same fixed Plummer potential is used for the LMC.  We search for the best few models which can explain the observed properties of the MS, LA, Bridge, and SMC at the present day.

Almost all previous investigations of the formation of the MS adopted a bulgeless disk (``pure disk'')  for the SMC (e.g., GN96, YN03, C06, B10, and DB11a,b), and we accordingly adopt a pure disk model as a first step.  However, recent observations have suggested that the older stellar populations of the SMC are supported by their velocity dispersion rather than rotation, implying that the SMC may possess a spheroid population of old stars (Harris \& Zartisky 2006).  Considering this observational result, our final investigation involves ``disk-plus-spheroid" models in which the baryonic component of the SMC is composed of a disk and a central spheroid.  We follow a two-step process: first, we run medium resolution N-body simulations representing the SMC by a self-gravitating disk, and we thereby identify the best tidal models from the previous set of test particle simulations; and second, we represent the SMC by a multi-component disk-plus-spheroid system and run a final set of high resolution simulations for our adopted model.  { In the present work, we ignore drag forces induced by the hot halo of the Milky Way (e.g., Diaz \& Bekki 2011b) and focus instead on the essential gravitational dynamics of the system.}

Numerical computations were carried out at the University of Western Australia on (i) the latest version of GRAPE (GRavity PipE, GRAPE-DR), which is a special-purpose
computer for gravitational dynamics (Sugimoto et al. 1990), and (ii) a Core i5 desktop computer system with a GPU card (NVIDIA GTX580) implementing the CUDA G5/G6 software package for calculations of gravitational dynamics.

The time integration of the equations of motion is performed by using a second-order leapfrog method with a time step interval of $\sim 0.02 t_{\rm dyn}$, where $t_{\rm dyn}$ is the dynamical time scale of the SMC.  To save on computational cost, we do not simulate over the full 5 Gyr of the computed orbit.  Instead, we begin our simulations 1 Gyr or so prior to the first strong interaction between the SMC and LMC.  For our adopted model, the simulated time window was 3.37 Gyr { (see Appendix for discussion)}.  The total number of particles for the first set of models (medium resolution, pure disk models) is $N$=200,000, split evenly between the disk and halo components.  The total number of particles in the second set of models (high resolution, disk-plus-spheroid models) is $N=10^6$, where $N_{\rm dm}$=500,000 for the dark matter halo, $N_{\rm d}$=400,000 for the disk, and $N_{\rm sph}$=100,000 for the spheroid.  The gravitational softening lengths $\epsilon$ for all components are set to be 72 pc.

\subsubsection{Pure Disk Models} \label{sec:mod-nbody-pure}

In this intermediate step we represent the SMC as a bulgeless disk galaxy embedded in a massive dark matter halo.  The total mass and the size of the dark matter halo (disk) is $M_{\rm dm, SMC}$ ($M_{\rm d, SMC}$) and $R_{\rm dm, SMC}$ ($R_{\rm d, SMC}$), respectively.  We adopt an NFW profile for the density distribution of the halo where the scale length $r_{\rm s, SMC}$ is set to be $0.5R_{\rm d, SMC}$.  { The mass ratio of dark matter to total SMC mass is fixed at $f_{\rm dm, SMC}=0.5$, which is consistent with the observed HI rotation curve for the SMC (e.g., Bekki \& Stanimirovi\'c 2009).  Although we did not explore other values for $f_{\rm dm, SMC}$, we should point out that a wide range of values could be consistent with the HI rotation curve depending on stellar mass-to-light ratios.}

The radial ($R$) density profile of the SMC disk is assumed to be proportional to $\exp (-R/R_{0, SMC})$ with scale length $R_{0, SMC} = 0.2R_{\rm d, SMC}$, and the vertical ($Z$) density profile is assumed to be proportional to ${\rm sech}^2 (Z/Z_{0, SMC})$ with scale length $Z_{0, SMC} = 0.04R_{\rm d, SMC}$.  In addition to the rotational velocity induced by the gravitational field of the disk and halo, the disk is assigned both radial and azimuthal velocity dispersions according to the epicyclic theory with Toomre's parameter $Q$ = 1.5.  The vertical velocity dispersion at a given radius is set to be 0.5 times as large as the radial velocity dispersion at that point.

{ Following the methodology of GN96}, the spin of the SMC disk is specified by two angles: ${\theta}_{\rm d}$, which is the angle between the $Z$-axis { of the Milky Way} and the angular momentum vector of the SMC disk, and ${\phi}_{\rm d}$, which is the azimuthal angle.  In particular, ${\phi}_{\rm d}$ is measured from the $X$-axis to the projection of the angular momentum vector onto the $X$-$Y$ plane of the Milky Way { (see Fig~1 of GN96 for schematic)}.  We investigated the full range of values for ${\theta}_{\rm d}$ and ${\phi}_{\rm d}$ for each of our best models, { and we discuss the range of preferred values in the Appendix}.  After analyzing the full set of pure disk models, we ended up with a final set of two best models, our \emph{adopted} model and an \emph{alternate} model.  Although these two models are remarkably similar in terms of how well they reproduce the MS, LA, and Bridge, { we will focus mostly on the adopted model and leave the alternate model to the Appendix.}  Table~\ref{tab:param1} gives the final set of parameters, and Fig~\ref{fig:sep} gives the orbital separations for the adopted model over 5 Gyr.

\subsubsection{Disk-plus-Spheroid Models} \label{sec:mod-nbody-smc}

Whereas the numerical investigations up to this point were solely for the purpose of finding the best model, the N-body simulations described here are presented as our main results in sections~\ref{sec:disk} and~\ref{sec:sph}.  We represent the SMC as a multi-component system composed of an NFW dark matter halo, an exponential disk, and a central spheroid, similar to the models of Bekki \& Chiba (2009).  The N-body prescriptions for the dark matter halo and the exponential disk are the same as those described in the pure disk model.  The spheroid is represented by a Plummer model with mass $M_{\rm sph, SMC}$, scale length $a_{\rm sph, SMC}$, and truncation radius $R_{\rm sph, SMC}$.  We make the simple assumption that the spheroid pertains to an old population of ``stars" and that the disk (particularly its outer parts) pertains to ``gas."  We therefore investigate whether the particles stripped from the disk are able to reproduce the observed properties of the MS. 

Although Nidever et al. (2011) have recently investigated the outer structure of metal-poor giant branch stars in the SMC, it remains observationally unclear how stellar populations of different ages and metallicities distribute within the tidal radius of the SMC.  We therefore assume that $R_{\rm sph, SMC}$ and $a_{\rm sph, SMC}$ are free parameters.  Likewise, we consider the mass ratio of the spheroid to the disk $f_{\rm sph, SMC}$ to be a freely adjustable parameter.  In our investigation of the formation processes of the MS, we find that reasonable variations in the mass ratio and scale length of the spheroid do not strongly affect its tidal evolution, and we accordingly adopt $f_{\rm sph, SMC}=0.2$ and $a_{\rm sph, SMC}=0.2R_{\rm sph, SMC}$.  However, changes in the truncation size of the spheroid strongly alter the tidal evolution, as described in section~\ref{sec:sph}.  Accordingly, we present the results of three high resolution models where the size of the spheroid is varied: an extended model (``Model 1"; $R_{\rm sph, SMC}=7.5$ kpc), an intermediate model (``Model 2"; $R_{\rm sph, SMC}=5$ kpc), and a compact model (``Model 3"; $R_{\rm sph, SMC}=2.5$ kpc).  Parameter choices for our best disk-plus-spheroid models are summarized in Table~\ref{tab:param2}.

Despite the simplicity of the present study (e.g., neglect of gas dynamics, star formation, chemical evolution), this work is nevertheless an important step toward constructing a self-consistent model for the formation of the MS and the evolution of the Magellanic system as a whole.  First, we have run a large number of models and found a promising set of parameters (including a set of proper motions for the MCs) which may well inform future modeling.  And second, as detailed in the following sections, we are able to explain many of the properties of the Magellanic system and MS using collisionless dynamics alone.  Where the model is unsuccessful, we consider the inclusion of gas dynamics to be a salient option for future improvements (see Discussion section~\ref{sec:dx-la}).  In particular, insights will surely be provided by modeling the hydrodynamical interaction of MS gas clumps with the hot halo of the Milky Way (e.g., Bland-Hawthorn et al. 2007).  
 

\begin{figure*}[t]\centering
\includegraphics[trim=20 0 0 0, scale=0.4]{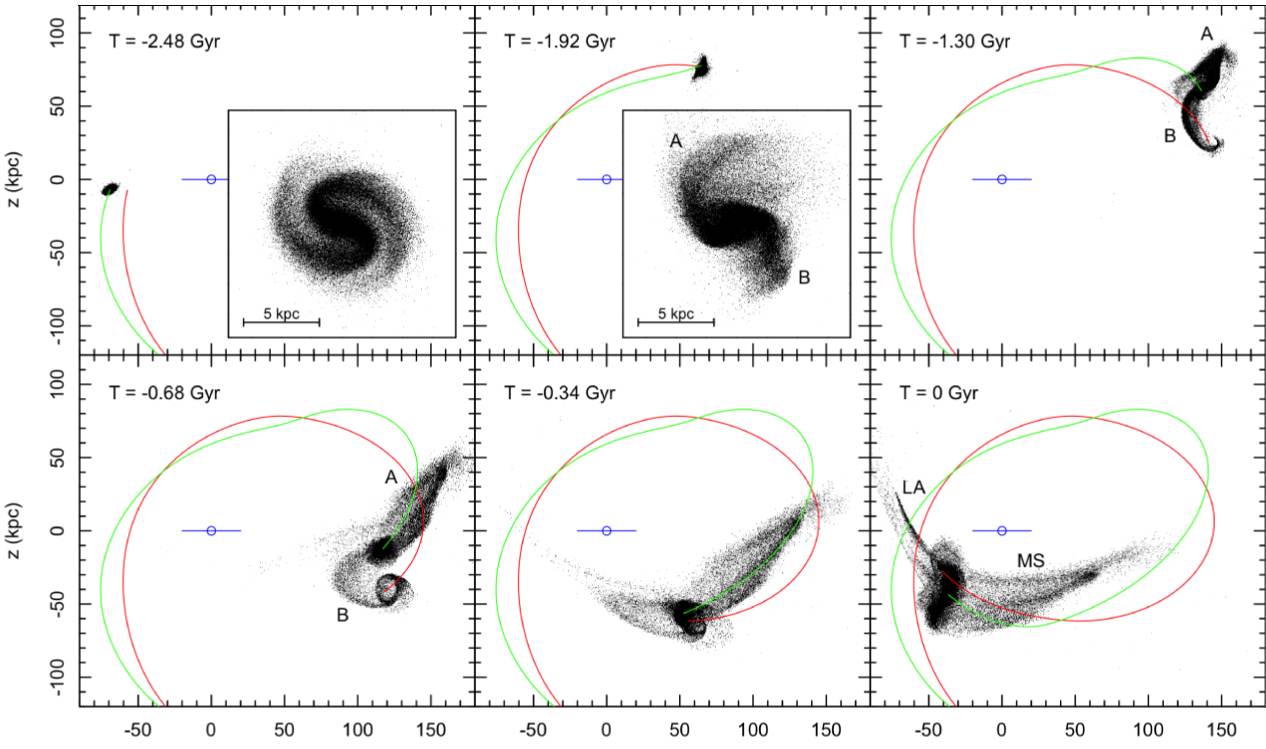}
\caption{The tidal evolution of the SMC disk. Each panel shows the distribution of SMC disk particles in the y-z plane at a specific time.  Also displayed are the orbit trails of the SMC (green) and LMC (red).  The Milky Way is shown schematically as an edge-on disk with radius 20 kpc, and the position of the Sun at (-8.5,~0,~0) kpc is indicated by the blue circle.  Because of the chosen projection in the y-z plane, the location of the Sun appears to coincide with the Galactic center (0,~0,~0).  In each of the first two panels an inset is provided which shows the face-on state of the SMC disk.  At $t=-2.48$ Gyr (top left panel and inset) the SMC disk has attained spiral structure due to weak tidal interaction with the LMC and Milky Way.  At $t=-1.92$ Gyr (top middle panel and inset) the SMC disk is disrupted by a close passage of the LMC, causing two tidal arms (A and B) to be stripped away.  At $t=-1.30$ Gyr (top right panel) arm A lags behind the SMC orbit while arm B is engulfed by the LMC.  Eventually, arm B forms a ring structure in the LMC ($t=-0.68$ Gyr, bottom left panel).  The leading and trailing debris elongate as the Milky Way pericenter is approached ($t=-0.34$ Gyr, bottom middle panel), and by $t=0$ Gyr (bottom right panel) the Magellanic Stream (MS) and Leading Arm (LA) are fully formed.  Not pictured is the formation of the Magellanic Bridge and Counter-Bridge, which happens at $t=-0.26$ Gyr as described in the text.
\label{fig:disk-time}}
\end{figure*}

\begin{figure*} \centering
\includegraphics[scale=0.6]{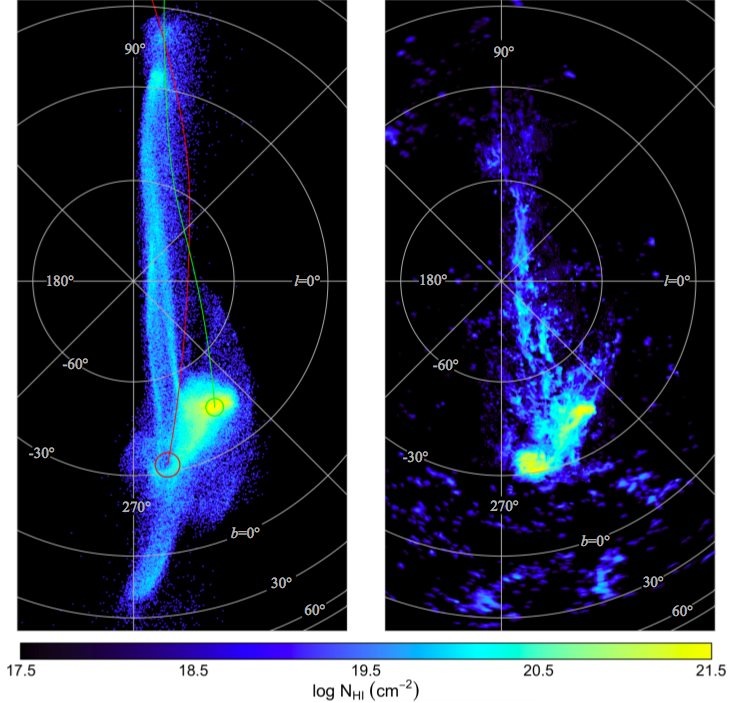}
\caption{(Left panel) The final distribution of disk particles projected onto the sky.  Column densities are computed as a smoothed intensity map of the simulated mass distribution.  The coordinates are galactic longitude $l$ (straight lines) and latitude $b$ (concentric circles) shown in a ZEA projection centered on the South Galactic Pole.  The orbit trails of the SMC (green) and LMC (red) are also shown, as are their current locations (circles).  (Right panel) The observed HI column densities of the Magellanic system, combined from data presented in Putman et al. (2003a) and Nidever et al. (2010).  The densest regions (yellow) are the SMC ($l=302.8^{\circ}$, $b=-44.6^{\circ}$) and LMC ($l=280.5^{\circ}$, $b=-32.5^{\circ}$).  The Magellanic Stream is the prominent trail of gas that splits into two filaments parallel to $l=270^{\circ}$ and $l=90^{\circ}$.  Other notable features include the Magellanic Bridge, which extends between the LMC and SMC, and the many branches of the Leading Arm at the bottom of the panel.  Column densities are represented on a logarithmic scale as shown at the base of the figure.
\label{fig:disk-coldens}}
 \end{figure*}

\begin{figure} \centering
\includegraphics[scale=0.45]{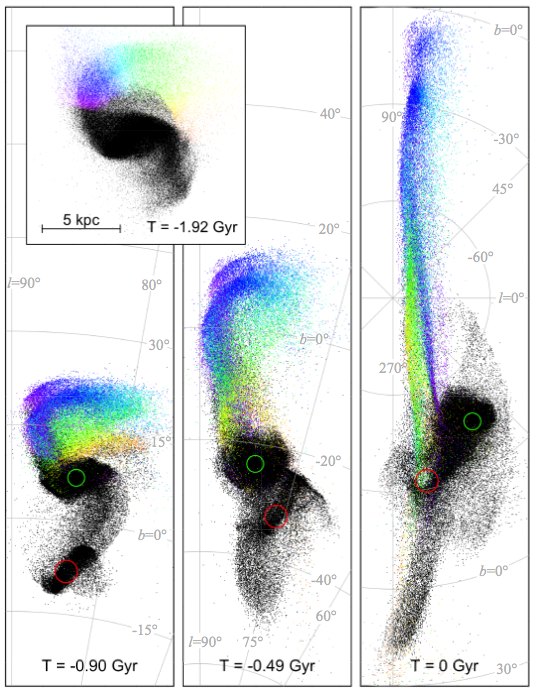}
\caption{The evolution of the Magellanic Stream, shown as a time series starting from the inset (top left, $t=-1.92$ Gyr) and proceeding to the left panel ($t=-0.90$ Gyr), then the middle ($t=-0.49$ Gyr), and finally the right ($t=0$ Gyr).  The coloring is chosen at $t=-1.92$ Gyr (inset), where the particles in tidal arm A (see Figure~\ref{fig:disk-time}) are colored according to their distance along the tidal arm, and all other particles are black.  This coloring is preserved at each time step as the tidal arm separates from the SMC (left and middle panels) and finally coincides with the present day Magellanic Stream (right panel).  See text for an analysis including a description of the dynamics.  The projection used in the right panel ($t=0$ Gyr) is an on-sky projection in galactic coordinates ($l$, $b$), and for ease of reference the same projection is used for the middle and left panels.  The SMC and LMC are represented as circles (green and red, respectively).  The inset is taken from Figure~\ref{fig:disk-time}, albeit without color.
\label{fig:disk-rain}}
 \end{figure}

\section{Results I - The SMC Disk} \label{sec:disk}

Here we describe our results concerning the SMC disk of Model 1.  See Tables~\ref{tab:param1} and~\ref{tab:param2} for parameter sets.

\subsection{Tidal Evolution of the SMC Disk} \label{sec:disk-time}

In our adopted model, the LMC and SMC have suffered only two close passages of one another, at $t=-1.97$ Gyr (the \emph{first} interaction) and $t=-0.26$ Gyr (the \emph{second} interaction) as seen in Fig~\ref{fig:sep}.  Previous to these encounters, the LMC and SMC were separated by large distances (up to $\sim120$ kpc at $t=-5$ Gyr) as they orbited independently within the virial radius of the Milky Way halo.  In this sense, the formation of a strong binary state is a recent phenomenon for the MCs, and this scenario provides a framework for understanding much of the dynamics.  For example, the strong tidal forces suffered by the SMC during the first interaction are responsible for stripping away much of its disk.  Figure~\ref{fig:disk-time} shows the time evolution of the SMC disk along with the orbital evolution of the MCs about the MW, and a snapshot of the disruption of the SMC is given in the top middle panel.

Previous to this epoch of strong interaction, the SMC is subject to a combination of weak tidal fields from the LMC and MW (e.g., at $t<-2$ Gyr, Fig~\ref{fig:sep}).  These weak tidal forces are sufficient to induce a bar and spiral arms within the SMC (Fig~\ref{fig:disk-time}, top left panel), but they are not strong enough to disrupt the disk.  In particular, even though the MW pericenter at $t=-2.3$ Gyr precipitates the formation of the strong LMC-SMC binary state, the MW pericenter itself does not actively participate in the removal of material from the SMC.  The disk remains intact up until the first close passage of the LMC at $t=-1.97$ Gyr, when the mutual separation between the MCs reaches a minimum of 6.0 kpc.  At this point, two tidal arms (``A" and ``B", top middle panel of Fig~\ref{fig:disk-time}) begin to separate from the main body of the SMC.  These two tidal arms follow quite different evolutionary paths: arm A falls into orbit behind the SMC and slowly elongates to form the Magellanic Stream, whereas arm B stretches ahead of the orbit and is largely engulfed by the LMC (Fig~\ref{fig:disk-time}, top right and bottom left panels).  As explored in the Discussion Sec~\ref{sec:dx-inlmc-gas}, this transfer of SMC disk material to the LMC would fuel star formation via the infall of metal-poor gas.  At $t=-0.68$ Gyr (Fig~\ref{fig:disk-time}, bottom left panel) the captured mass within the LMC resembles a polar ring, suggestive of the distribution of carbon stars in the present-day LMC (Kunkel et al. 1997a).

By $t=-0.34$ Gyr, the MW has accelerated a fraction of the stripped material into the leading direction of the orbit (Fig~\ref{fig:disk-time}, bottom middle panel).  Though not immediately clear from Fig~\ref{fig:disk-time}, the leading material is collected into two separate branches, one that originates in tidal arm A and the other in B.  In Sec~\ref{sec:disk-dkin} we will indicate that these branches are spatially and kinematically distinct.  At $t=-0.26$ Gyr, the LMC and SMC undergo their second strong interaction, reaching a minimum separation of 6.6 kpc.  This second close passage strips an additional two tidal arms from the SMC disk, but only \emph{one} of these structures corresponds to an observable HI feature of the Magellanic system, namely the Magellanic Bridge.  The other structure, which we call the ``Counter-Bridge," has perhaps eluded identification by extending almost directly behind the SMC along the line of sight.  The three-dimensional nature of the Bridge and Counter-Bridge is considered in Sec~\ref{sec:disk-bcb}.  At $t=0$ Gyr, the MCs have arrived to their present-day positions and velocities, and the simulated particle distribution has reached its final state.  The locations of the simulated Leading Arm and Magellanic Stream are noted in the bottom right panel of Fig~\ref{fig:disk-time}.

\subsection{Projected On-sky Distribution} \label{sec:disk-sky}

To make a meaningful comparison with observations, we must project the final distribution of particles against the sky and create a map of their density.  The left hand panel of Fig~\ref{fig:disk-coldens} gives the simulated column density distribution in a ZEA projection centered on the south galactic pole.  The coordinates are galactic longitude ($l$) and latitude ($b$).  The orbit trails of the SMC and LMC are provided (green and red, respectively), as are their current locations (circles).  The right hand panel gives the observed HI column density map of the Magellanic system, where column densities are plotted on the same logarithmic scale as the simulation data.  Although the observed LMC is rich in HI (right panel), our simulated column density map does not display a high density region at the location of the LMC.  This is simply due to our model assumptions, because the LMC is represented by a fixed potential rather than a live ensemble of particles.  { The spatial extent of HI in the SMC and Bridge appears to be larger in the simulation than in observation, which is also the case in the collisionless simulations of Connors et al. (2006).  This discrepancy could be mitigated by including gaseous pressure from the Milky Way's hot halo, which would confine the gas.}

Judging from Fig~\ref{fig:disk-coldens}, the simulation provides a good reconstruction of the Bridge and MS in terms of on-sky location and morphological structure.  The leading material, however, does not provide a good match to the observed location of the LA.  We argue in Sec~\ref{sec:dx-la} that the inclusion of ram pressure from the Milky Way's hot halo may be able to resolve this discrepancy.  The tip of the MS also poses a problem, because the simulation over-predicts the density as compared to observations, particularly for $b>-40^{\circ}$.  It would appear from Fig~\ref{fig:disk-coldens} that the observed MS does not even extend into this distant region, but Nidever et al. (2010) have shown that MS gas does indeed exist here, albeit at low column densities.  As compared to the observations of Nidever et al. (2010), our simulation correctly predicts the total length of the { stripped Magellanic gas $(\sim200^{\circ}$ from MS tip to the end of the LA)}, and it also correctly predicts the kinematical structure of the MS tip (Sec~\ref{sec:disk-dkin}).

The triangle-shaped structure at $(l$, $b)=(315^{\circ}$, $-50^{\circ})$ (Fig~\ref{fig:disk-coldens}, right panel) is reproduced reasonably well in our model, despite having a slightly different orientation.  The structure has not yet been given a unique identity in either the observational or theoretical literature: no simulation has yet addressed its formation, and observers have associated it with the gas clouds at the base of the MS (e.g., Putman et al. 2003a; Bruns et al. 2005, who use the collective label ``Interface Region").  We find that the origin of this structure is quite different to that of the MS, and we therefore propose it be named to reflect its unique identity.  In the present work we will use the name ``Magellanic Horn," owing to its tapered appearance.  Whereas the MS is a product of the elongation of tidal arm A, the Horn is a remnant of tidal arm B.  When portions of tidal arm B are accreted onto the LMC, much of the additional material is left in orbit around the LMC periphery.  Some of this peripheral material is slung away as the LMC and SMC plunge toward one another at $t=-0.26$ Gyr, becoming unbound from the MCs.  The material that is left behind in the trailing direction of the orbit gradually elongates to form the Horn.

Perhaps the most compelling aspect of the simulation is the reproduction of MS bifurcation.  As seen in the right panel of Fig~\ref{fig:disk-coldens}, the base of the MS bifurcates into two parallel and distinct filaments, and the location of these filaments is well reproduced in the simulation.  An important observational feature of the filaments is that they appear to cross at several locations along the MS (Putman et al. 2003a).  The location of the first crossing point at $(l$, $b)\approx(45^{\circ}$, $-80^{\circ})$ is clearly exhibited in the simulated MS, providing a convincing morphological match.  Considering the strong reproduction of the MS filaments, it will be instructive to describe their origin in detail.  We do so in the following section, { and we compare with the findings of Nidever et al. (2008) in Sec~\ref{sec:dx-compare}}.

\subsection{The Origin of MS Bifurcation} \label{sec:disk-bif}

Figure~\ref{fig:disk-rain} color-codes the time-evolution of the MS.  To construct the color-code, we first fit a logarithmic spiral to tidal arm A at $t=-1.92$ Gyr (Fig~\ref{fig:disk-time}, inset of top middle panel), and then we assign color to the particles based on their projected distance along the arm (Fig~\ref{fig:disk-rain}, inset).  The colors run across a portion of the HSV spectrum beginning at the base of tidal arm A (purple and blue) continuing to its middle (cyan and green) and ending at its tip (yellow and orange).  The particles retain their color throughout the evolution, allowing us to trace the origin of structures as the MS is formed.

As seen in the right panel ($t=0$ Gyr) of Fig~\ref{fig:disk-rain}, the two filaments of the MS originate from opposite ends of tidal arm A.  The right filament originates from the base of arm A (purple and dark blue) whereas the left filament comes from its middle and end (green and yellow, respectively).  In addition, the colors that fall in between these extremes are found in the MS tip (blue, cyan).  Accordingly, we can think of the MS as being a wrapping of tidal arm A: (i) the base of arm A becomes the base of the right MS filament, (ii) arm A is traced upward through the right filament toward the MS tip, and (iii) the middle of arm A is traced downward through the left MS filament.  Though it appears faint in Fig~\ref{fig:disk-rain}, the tip of arm A (orange) continues to trace downward through the left filament into the leading region.  The structure of the leading material is made more clear in Sec~\ref{sec:disk-dkin}.

More than simply tracing up and back down the MS, the material of tidal arm A also \emph{crosses} within the body of the MS.  Consider the left panel of Fig~\ref{fig:disk-rain} ($t=-0.90$ Gyr), in which the material from the base of arm A (purple and blue) is to the \emph{left} of that from the end of arm A (green and yellow).  In the middle panel ($t=-0.49$ Gyr), the material from the base \emph{overlaps} with that from the end of arm A, and finally in the right panel ($t=0$ Gyr), the material from the base is to the \emph{right} of that from the end.  In other words, the relative on-sky orientation of the MS filaments has reversed within the last $\sim$1 Gyr.  This scenario provides a physical interpretation for the crossing point at $(l$, $b)\approx(45^{\circ}$, $-80^{\circ})$: it is a region of physical overlap, similar to the crossover point of a ribbon that has been folded across itself.

But what dynamical processes are responsible?  One important point of emphasis is that the tip of the MS (blue) does \emph{not} come from the tip of tidal arm A (yellow-orange).  This idea runs counter to the expectation from two-body tidal stripping models (e.g., Toomre \& Toomre 1972; Besla et al. 2010) in which the tips of tidal tails elongate away from their parent galaxies.  The complex evolution of tidal arm A underlines the \emph{three-body} nature of the formation of the MS.  Tidal arm A is removed from the SMC disk due to a close passage of the LMC, but once it becomes unbound, its subsequent evolution is dominated by the MW.  In particular, the approach to MW pericenter over the past $\sim$1 Gyr is responsible for elongating tidal arm A and creating an identifiable MS tip, as seen in the time progression of Fig~\ref{fig:disk-rain}.  Because they are created by distinct dynamical processes, there is no reason to expect that the tip of the MS should coincide with the tip of tidal arm A.

The MW also accelerates much of the unbound material of tidal arm A into the leading direction of the orbit.  In the left panel of Fig~\ref{fig:disk-rain} ($t=-0.90$ Gyr), the entirety of arm A is on the trailing side of the orbit, but as MW pericenter is approached (middle and right panels), the tip of arm A (yellow-orange) is accelerated to the leading side.  Notice that on its path to the leading side (as seen by comparing the three main panels of Fig~\ref{fig:disk-rain}), the tip of arm A (yellow-orange) must cross from the right side to the left side of the purple/blue filament.  The MW coaxes the adjoining material of arm A (i.e., the green/yellow filament) to follow a similar evolutionary path, causing it also to cross over.  The material from the base of arm A (i.e., the purple/blue filament) is perhaps less susceptible to this effect because it originates closer to the SMC at the \emph{base} of tidal arm A, and it is therefore more strongly anchored in place.  

\begin{figure*} \centering
\includegraphics[scale=0.5]{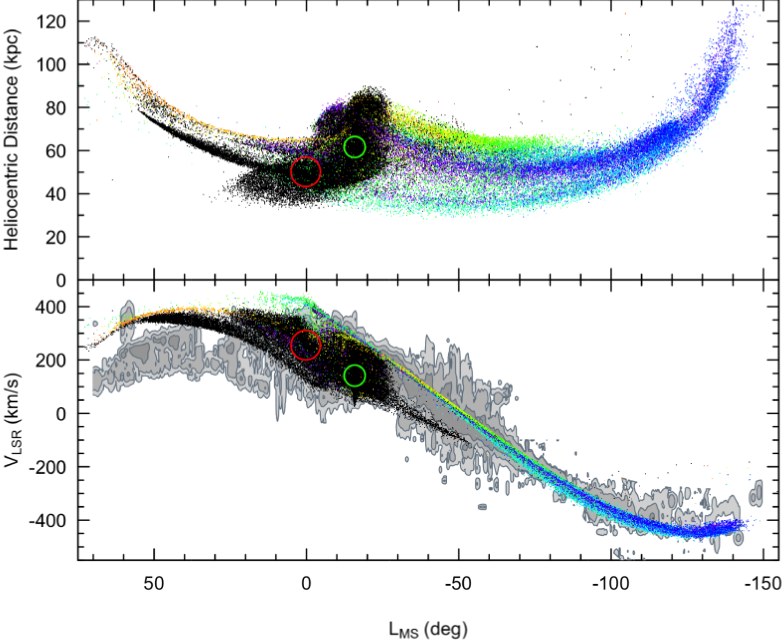}
\caption{The heliocentric distances (top panel) and $v_{\rm LSR}$ velocities (bottom panel) of the disk particles.  The horizontal axis is Magellanic longitude L$_{\rm MS}$, which runs parallel to the Magellanic Stream and is defined by Nidever et al. (2008).  The coloring of the particles is maintained from Figure~\ref{fig:disk-rain}.  In each panel, the Leading Arm is to the left of the SMC (green circle) and LMC (red circle), whereas the Magellanic Stream is to the right.  In the bottom panel, observational data from Nidever et al. (2010) is represented by grey filled contours (at 1, 10$^{-1.0}$, and 10$^{-2.5}$ Kelvin deg).
\label{fig:disk-rhovlsr}}
 \end{figure*}

\begin{figure} \centering
\includegraphics[trim=10 0 0 0, scale=0.35]{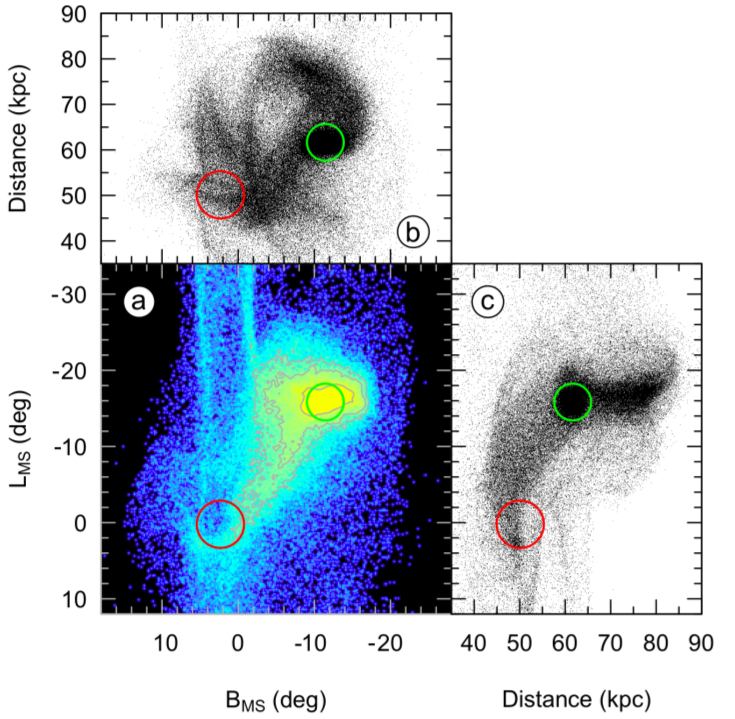}
\caption{The Magellanic Bridge and Counter-Bridge.  (a) A close-up of the column density map of Figure~\ref{fig:disk-coldens} using the same color scheme.  The panel is centered on the Magellanic Bridge, which extends between the SMC (green circle) and LMC (red circle).  The on-sky coordinates are Magellanic longitude L$_{\rm MS}$ and latitude B$_{\rm MS}$ as defined by Nidever et al. (2008).  Contours are drawn at the 0.96, 0.98, and 0.995 quantiles, outlining the Bridge and its tidal complement the Counter-Bridge, which is located at $({\rm L}_{\rm MS}$, ${\rm B}_{\rm MS}) \approx (-16^{\circ}, ~-6^{\circ})$.  The heliocentric distances of the simulation particles are plotted against (b) Magellanic latitude ${\rm B}_{\rm MS}$, and (c) Magellanic longitude ${\rm L}_{\rm MS}$.  The coordinates of the SMC and LMC are shown as green and red circles, respectively.  The presence of the Counter-Bridge becomes more obvious in panel (b) where it is located at top right, between ${\rm B}_{\rm MS} \approx -4^{\circ}$ and $-16^{\circ}$, and in panel (c) where it is located at right, between ${\rm L}_{\rm MS} \approx-14^{\circ}$ and $-20^{\circ}$.  As seen in both panels, the Counter-Bridge extends up to 20 kpc away from the SMC largely along the line of sight.
\label{fig:disk-bcb}}
 \end{figure}

\begin{figure} \centering
\includegraphics[trim=20 0 0 0, scale=0.35]{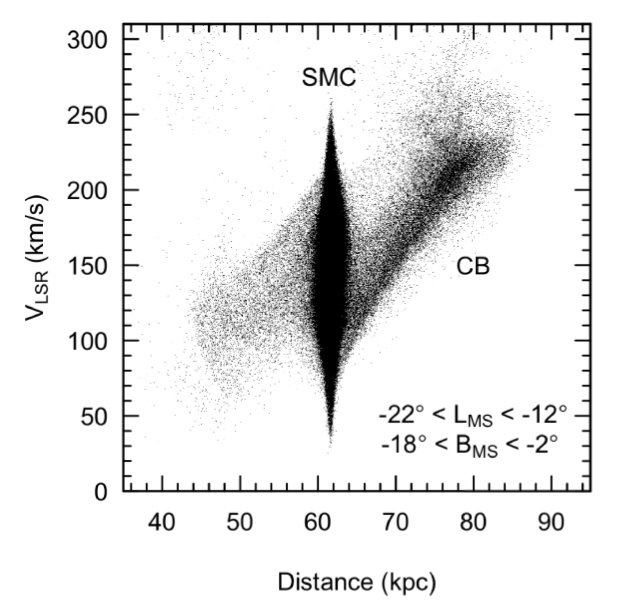}
\caption{Velocity $v_{\rm LSR}$ plotted as a function of heliocentric distance $D$ for particles originating within the disk of the SMC.  The only particles plotted here are roughly coincident with the SMC as projected onto the sky ($-22^{\circ}< {\rm L}_{\rm MS} <-12^{\circ}$, $-18^{\circ}<  {\rm B}_{\rm MS} <-2^{\circ}$).  The location of the SMC (59 kpc $< D <$ 65 kpc) and Counter-Bridge ($D >$ 65 kpc; ``CB") are labeled.
\label{fig:new-cb}}
 \end{figure}

\begin{figure} \centering
\includegraphics[trim=20 0 0 0, scale=0.3]{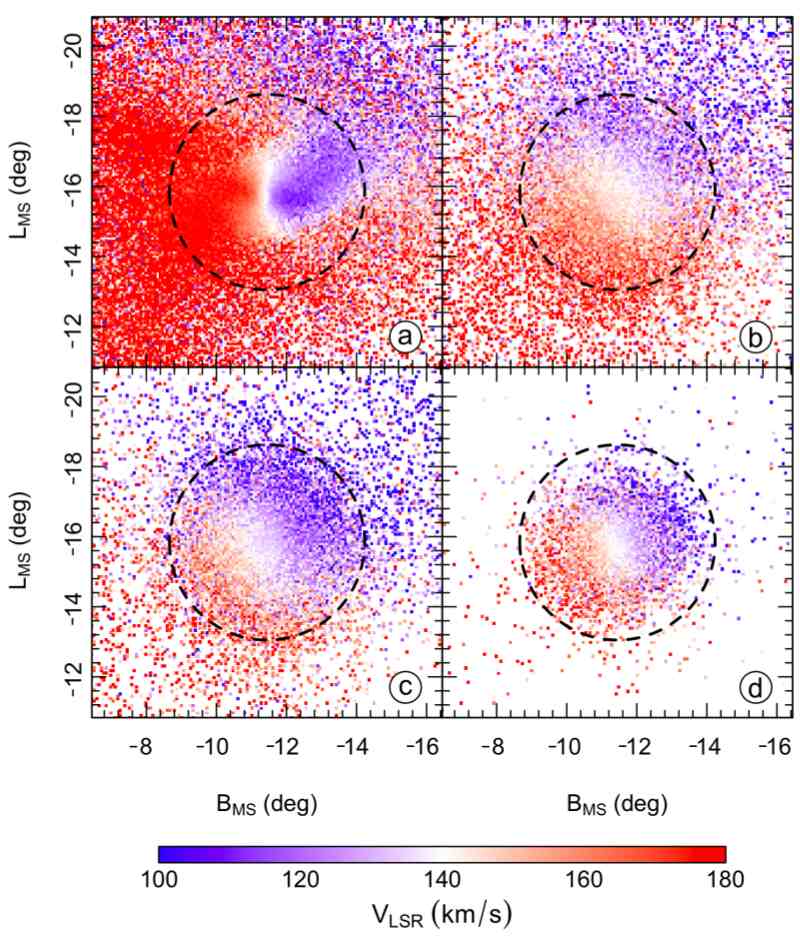}
\caption{The $v_{\rm LSR}$ velocity field of (a) the SMC disk of model 1, and (b-d) the SMC spheroid of models 1-3, respectively.  In each panel, the on-sky coordinates are Magellanic longitude L$_{\rm MS}$ and latitude B$_{\rm MS}$ as defined by Nidever et al. (2008).  The dashed circle has a radius of 3 kpc and is centered on the SMC center of mass.  The velocity range has a lower bound of 100 km s$^{-1}$ (blue), an upper bound of 180 km s$^{-1}$ (red), and is centered on the $v_{\rm LSR}$ velocity of the SMC (white).  Only the disk (a) exhibits a large rotational amplitude, because the velocity gradient for the spheroids (b-d) is due to tidal effects (see text).
\label{fig:disk-vel}}
 \end{figure}

\begin{figure*}[t]\centering
\includegraphics[trim=20 0 0 0, scale=0.4]{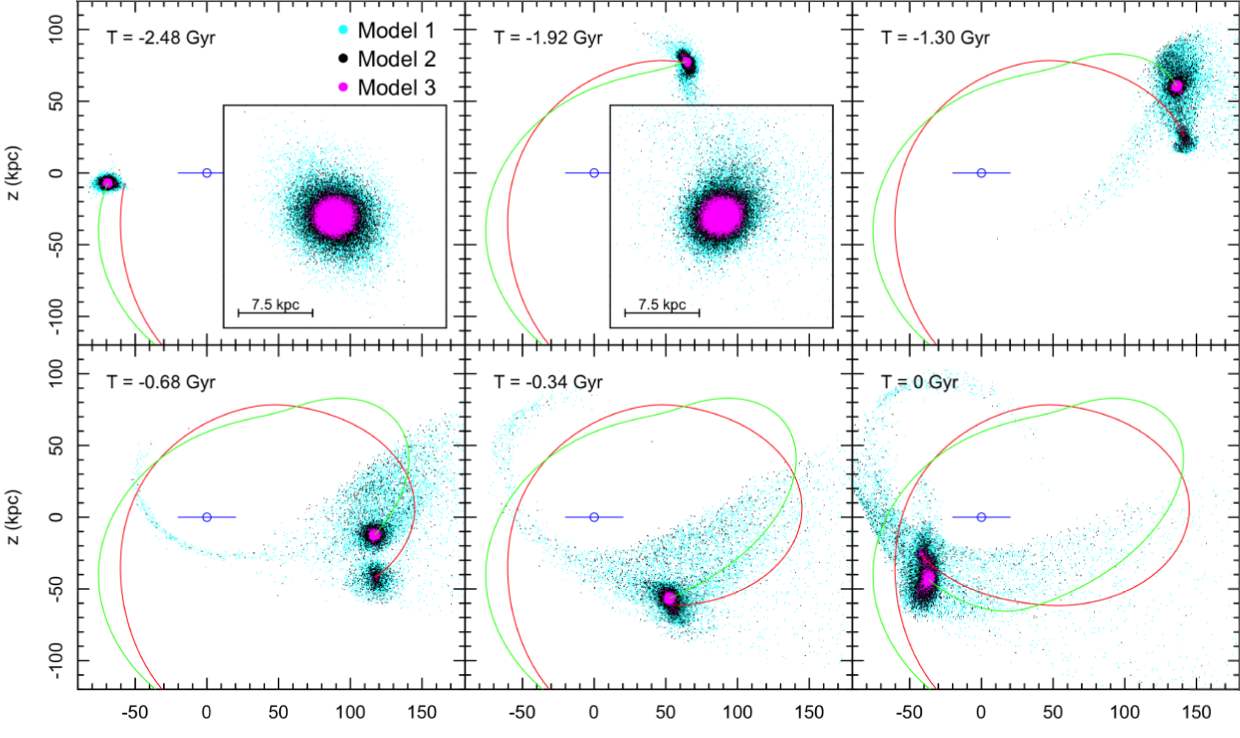}
\caption{The tidal evolution of the SMC spheroid. Each panel shows the distribution of SMC spheroid particles of models 1 (cyan), 2 (black), and 3 (magenta) in the y-z plane at a specific time.  Also displayed are the orbit trails of the SMC (green) and LMC (red).  The Milky Way is shown schematically as an edge-on disk with radius 20 kpc, and the position of the Sun at (-8.5,~0,~0) kpc is indicated by the blue circle.  Because of the chosen projection in the y-z plane, the location of the Sun appears to coincide with the Galactic center (0,~0,~0).  In each of the first two panels an inset is provided which shows the face-on state of the SMC spheroid.  At $t=-2.48$ Gyr (top left panel and inset) the SMC spheroid is only weakly distressed by tidal forces, and the key differences between model 1 (extended), model 2 (intermediate), and model 3 (compact) are clear.  The close passage of the LMC at $t=-1.92$ Gyr (top middle panel and inset) causes  a number of particles to be stripped away, particularly for models 1 and 2.  A portion of these stripped particles are engulfed by the LMC ($t=-1.30$ Gyr, top right), while others orbit freely in leading and trailing streams ($t=-0.68$ Gyr, bottom left).  A second close passage at $t=-0.34$ Gyr (bottom middle) pulls more spheroid particles from the SMC.  At the present day ($t=0$ Gyr, bottom right) stellar analogues for the Magellanic Bridge, Leading Arm, and Magellanic Stream have been formed.\\
\label{fig:spher-time}}
\end{figure*}

\begin{figure} \centering
\includegraphics[trim=0 0 0 20, scale=0.45]{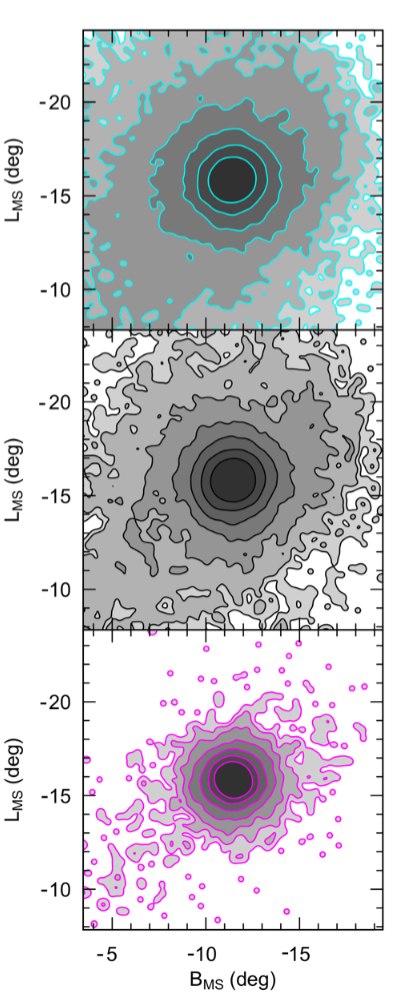}
\caption{Close-up of the final SMC spheroid distribution projected onto the sky for model 1 (top panel, cyan contours), model 2 (middle, black contours), and model 3 (bottom, magenta contours).  Contours are drawn at projected particle counts of 10$^x$ where $x=$(0.5, 1.0, 1.5, 2.0, 2.5, 3.0, 3.5), and the filled regions range from light grey to dark grey as $x$ increases.  The colors of the contours merely indicate the identity of each model (i.e., the contours represent the same densities, regardless of color).  The on-sky coordinates are Magellanic longitude L$_{\rm MS}$ and latitude B$_{\rm MS}$ as defined by Nidever et al. (2008).
\label{fig:new-zoom}}
 \end{figure}

\begin{figure} \centering
\includegraphics[trim=10 0 0 0, scale=0.34]{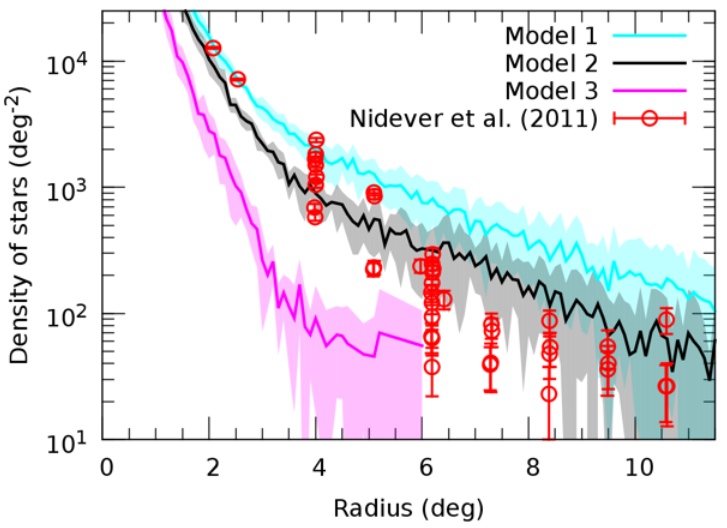}
\caption{The on-sky radial density profiles of the SMC spheroid for models 1 (cyan), 2 (black), and 3 (magenta) at the present day $t=0$.  To construct the profiles for each model, the spheroid particles are binned according to on-sky radius (bin size 0.1$^{\circ}$) and azimuthal angle (bin size $51^{\circ}$).  We use the on-sky location of the SMC center of mass in each model as the origin $r=0$, and we require at least one particle on average in each bin.  The solid lines indicate the average density at each radius (i.e., annular ring in the sky) for each model, and the shaded region gives the standard deviation across the azimuthal bins.   Observational data (red circles and error bars) is taken from the red giant survey of Nidever et al. (2011).  In order to facilitate comparison, the density profiles of the models have been rescaled such that the density of model 1 matches the Nidever et al. (2011) datapoint at $r\approx2^{\circ}$ (see text for details).
\label{fig:break}}
 \end{figure}

\subsection{Distance and Kinematics}  \label{sec:disk-dkin}

Fig~\ref{fig:disk-rhovlsr} gives the distances and $v_{\rm LSR}$ radial velocities of the simulation particles, and the color scheme of Fig~\ref{fig:disk-rain} has been retained in order to permit the identification of various structures.  The MS (colored) and Horn (black) are found on the trailing side of the orbit ($L_{MS}<-20^{\circ}$; right hand side of Fig~\ref{fig:disk-rhovlsr}), and the leading material is found on the leading side of the orbit ($L_{MS}>10^{\circ}$; left hand side of Fig~\ref{fig:disk-rhovlsr}).  It is clear that the leading material has two distinct branches, one that is sourced from tidal arm A (i.e., from its tip, colored yellow-orange as described in Sec~\ref{sec:disk-bif}), and the other from tidal arm B (colored black).  A comparison of the top and bottom panels of Fig~\ref{fig:disk-rhovlsr} shows that these two branches are both kinematically and spatially distinct, despite being overlapped in the on-sky map of Fig~\ref{fig:disk-coldens}. { The two branches are not separated on the sky simply because of a projection effect.}

The observed HI kinematics of the Magellanic system are given as grey contours in the background of Fig~\ref{fig:disk-rhovlsr}, lower panel (Nidever et al. 2010).  The correspondence between simulation and observation is good for the MS.  The main body of the MS correctly predicts the observed radial velocity profile, and the two MS filaments possess an essentially identical (i.e., overlapping) kinematical structure.  The tip of the MS shows a distinct velocity inflection that is echoed in the observations of Nidever et al. (2010).  This fact provides a good justification for the material at the tip of the simulated MS: even though it has the wrong density (see Fig~\ref{fig:disk-coldens}), it has the correct extent and the correct kinematics.  The radial velocity profile of the Horn is slightly offset to that of the MS, but this is to be expected because the Horn originates in tidal arm B along a very different evolutionary track (see Sec~\ref{sec:disk-sky}) as compared to the MS, which originates in arm A.

The correspondence between simulation and observation is comparatively poor for the LA.  The shape of the velcocity profile is generally correct for both branches of leading material, but it is offset by $\sim150$ km s$^{-1}$ from observations.  The heliocentric distances (Fig~\ref{fig:disk-rhovlsr}, upper panel) also suggest a discrepancy for the LA region.  Whereas the simulation predicts that the LA should lie at distances of $\sim50$ kpc to $\sim100$ kpc, McClure-Griffiths et al. (2008) have provided a much smaller distance of $\sim$20 kpc.  The uncertainty of this data point is considerable, because it is a kinematic distance derived from a supposed interaction between the LA and the HI disk of the MW.  Nevertheless, the large discrepancy in distances for the LA, coupled with the failure to correctly predict the kinematics and on-sky location, suggests that we are missing an essential ingredient in the dynamics.  As discussed in Sec~\ref{sec:dx-la}, we consider this ingredient to be ram pressure from the MW hot halo.

Observational data does not appear in the upper panel of Fig~\ref{fig:disk-rhovlsr} because distances to HVCs cannot be directly measured, barring the one data point of McClure-Griffiths et al. (2008) for the LA.  At best, only upper and lower bounds can be obtained by carefully considering foreground/background absorption spectra (e.g., Wakker 2001), but this has not been done extensively for the MS and LA, whose distance profiles are completely unknown (however, see Jin \& Lynden-Bell 2008).  Accordingly, it cannot yet be verified if the MS has a large radial extent as predicted in the simulation (Fig~\ref{fig:disk-rhovlsr}, upper panel).  This large radial spread is quite curious, especially when we consider that the on-sky projection of the MS is filamentary.  It may be more appropriate to describe the MS as radially extended \emph{sheets} rather than confined filaments.  The responsible dynamical process is simple: at the same time that the MW elongates the MS on the sky over the past $\sim$1 Gyr, radial tidal forces from the MW cause elongation along the line of sight.  These processes cannot be decoupled, but perhaps the elongation is mitigated by hydrodynamical interactions such as cloud confinement and ablation (e.g., Bland-Hawthorn et al. 2007).

\subsection{Magellanic Bridge and Counter-Bridge} \label{sec:disk-bcb}

It is clear that the first interaction between the MCs (at $t=-1.97$ Gyr) removes two tidal structures from the SMC disk (e.g., Fig~\ref{fig:disk-time}), but it is less clear that an additional two structures are removed during the second strong interaction (at $t=-0.26$ Gyr).  At least one tidal structure obviously forms, because it fills the region in between the SMC and LMC in good agreement with the observed Magellanic Bridge.  The other structure is less obvious, however, and here we call it the ``Counter-Bridge".  Panel (a) of Figure~\ref{fig:disk-bcb} shows a close-up view of the simulated column density map of Fig~\ref{fig:disk-coldens} centered on the Magellanic Bridge.  In this panel, the existence of the Counter-Bridge is traced only by a small region of high density around $-L_{MS} \approx 16^{\circ}$, $-B_{MS} \approx 6^{\circ}$.  Panels (b) and (c) reveal the three-dimensional structure of the region by plotting the heliocentric distances of the simulation particles along each spatial direction.  These panels reveal a dense and clearly defined tidal structure, i.e., the Counter-Brige, extending away from the SMC toward distances of up to $d\approx85$ kpc.

Despite the $\sim20$ kpc extent of the Counter-Bridge, it remains propitiously aligned with the SMC along the line of sight and is therefore hidden in the on-sky column density distribution.  Perhaps for this reason alone the Counter-Bridge has eluded observational identification.  However, a ``loop" of HI on the northeast edge of the SMC may betray the existence of the Counter-Bridge, as suggested by Muller \& Bekki (2007).  Another possible observational analogue of the Counter-Bridge is the large line-of-sight depth of the SMC, which has been measured to be $\sim6$ to $\sim12$ kpc (Welch et al. 1987; Crowl et al. 2001; Lah et al. 2005), $\sim12$ to $\sim16$ kpc (Hatzidimitriou \& Hawkins 1989; Gardiner \& Hawkins 1991; Subramanian \& Subramaniam 2012), or up to $\sim20$ kpc (Mathewson et al. 1988; Groenewegen 2000).  Despite the disagreement in these observational measurements (see HZ06 for a clarifying discussion), it is clear than the SMC is significantly extended along the line of sight and particularly in the north-east quadrant.  The present model provides a natural mechanism to explain this structure, namely that a projection effect along the line of sight causes the bound stellar populations of the SMC to be confused with the unbound stars tracing the tidally extended Counter-Bridge.

{ The radial velocity signature of the Counter-Bridge is given in Fig~\ref{fig:new-cb}, which plots $v_{\rm LSR}$ against heliocentric distance $D$ for all particles that roughly overlap the SMC in the sky ($-22^{\circ}< {\rm L}_{\rm MS} <-12^{\circ}$, $-18^{\circ}<  {\rm B}_{\rm MS} <-2^{\circ}$).  There are three clear components, each residing at distinct distances: particles bound to the SMC (59 kpc $< D <$ 65 kpc), those in the Bridge ($D <$ 59 kpc), and those in the Counter-Bridge ($D >$ 65 kpc).  Because much of the Bridge is found elsewhere on the sky, the velocity profile for the Bridge is represented incompletely in Fig~\ref{fig:new-cb}.  On the other hand, the velocity profile of the Counter-Bridge is fully characterized in Fig~\ref{fig:new-cb}, which shows a linear rise in velocity with distance.  Despite this unique signature, the Counter-Bridge remains confused with the SMC in projection, because each exhibits the same range of radial velocities in roughly the same locations on the sky.

Accordingly, it may be difficult to observationally verify/refute the existence of the Counter-Bridge without obtaining distance measurements.  For this reason, HI observations are not as suitable to the task as stellar surveys.  The presence of young stars in the Bridge (e.g., Irwin et al. 1985) may indicate that young stars trace the Counter-Bridge as well, owing to the fact that the Bridge and Counter-Bridge are complementary tidal structures with similar origins.  The use of older stars as tracers of the Counter-Bridge may prove difficult because intermediate-age stars have yet to be observed in the Bridge (Harris 2007).  Regardless of the tracer population used, establishing a distance-velocity correlation akin to Fig~\ref{fig:new-cb} would provide a strong test of the existence of the Counter-Bridge.}

\subsection{SMC velocity profile} \label{sec:disk-smcv}
Panel (a) of Fig~\ref{fig:disk-vel} gives the $v_{\rm LSR}$ velocity field of the SMC disk at the present day.  There is a strong sign of rotation, consistent with the observed HI kinematics of the SMC (Stanimirovi\'c et al. 2004).  Rotation is not consistent, however, with the large population of red giant stars observed within the SMC (Harris \& Zaritsky 2006).  This old stellar population appears to be supported instead by its velocity dispersion, suggestive of a spheroidal morphology.  We are therefore motivated to investigate the tidal evolution of such a spheroid in the following section.
\\
\\

\section{Results II - The SMC Spheroid} \label{sec:sph}

Here we describe our results concerning the SMC spheroid of Models 1, 2, and 3 (see Tables~\ref{tab:param1} and~\ref{tab:param2} for parameter sets).  Because little is known about the evolution of the spheroid, we present multiple parameterizations as a comparative study.  The essential difference between these spheroids is their initial size ($R_{\rm sph, SMC}$): Model 1 is extended (=7.5 kpc), Model 2 is intermediate (=5.0 kpc), and Model 3 is compact (=2.5 kpc).

\subsection{Tidal evolution of the SMC spheroid} \label{sec:sph-time}
The SMC spheroid evolves under the same tidal forces as the disk (Sec~\ref{sec:disk-time}), and its evolution is of course coupled to that of the disk through the N-body dynamics.  Fig~\ref{fig:spher-time} shows the tidal evolution of each spheroid model along with the orbits of the MCs.  The particles belonging to models 1, 2, and 3 are colored cyan, black, and magenta, respectively.  We showed previously that the SMC disk was disrupted following the first strong interaction between the MCs at $t=-1.97$ Gyr, and we see from Fig~\ref{fig:spher-time} (top middle panel) that the same forces are responsible for disrupting the spheroid.  Only models 1 and 2 suffer significant mass loss, however, as the spheroid of model 3 is sufficiently compact to avoid widespread stripping.

Similar to the case of the disk, the unbound spheroid particles fall into orbits on both the leading and trailing side of the orbit, with a significant fraction being engulfed by the LMC (Fig~\ref{fig:spher-time}, top right and bottom left panels).  By $t=-0.34$ Gyr (Fig~\ref{fig:spher-time}, bottom middle panel), the MW has greatly elongated the leading and trailing streams.  The second close passage of the MCs at $t=-0.26$ Gyr successfully removes material from all three spheroid models, though model 3 once again suffers the least amount of stripping.  At $t=0$ Gyr (Fig~\ref{fig:spher-time}, bottom right panel), the MCs have arrived to their present-day positions and velocities, and the spheroid particles have reached their final distribution.  As can be seen, the leading and trailing streams contain particles from models 1 and 2 only, but particles from all three models are stripped into the region between the LMC and SMC (i.e., the Magellanic Bridge region).  Because the spheroid imitates an old stellar population (e.g., Harris \& Zaritsky 2006), the stripped structures can be taken as possible stellar analogues of the MS, LA, and Bridge.

\subsection{SMC Outer Halo} \label{sec:sph-halo}

{ In their recent survey of the stellar periphery of the SMC, Nidever et al. (2011; N11) discovered an extended halo component expanding to at least $r\sim11^{\circ}$ from the SMC.  Although it is observationally unclear whether this newfound component is bound or unbound, we can test if the tidal distortions of the SMC spheroid in our adopted model are able to reproduce this important new observation.  Fig~\ref{fig:new-zoom} gives a close-up view of the SMC for our three spheroid models.  Models 1 and 2 (top and middle panel, respectively) exhibit an extended halo population, but such a component is largely absent in model 3 (bottom panel), which remains dense and compact.  A bit of tidal distortion is evident for model 3 toward the Bridge region (i.e., to the lower left of the panel), but this cannot be described as an extended, azimuthally symmetric population as discovered by N11.

Further to their discovery, N11 show a distinct ``break point" in the radial density profile of red giants occurring $r\sim7^{\circ}$ from the SMC.  This break point (i.e., sharp change in slope) is indicative of tidal distortions (e.g., Mu\~noz et al. 2008) and should be reproduced in tidal models of the SMC.  In Fig~\ref{fig:break} we compare the N11 radial density data against the predictions of our three spheroid models, where lines give the average radial density of each model and shaded regions give the standard deviation of different azimuthal bins.  In other words, Fig~\ref{fig:break} provides not only the radial density but also its degree of symmetry in different directions of the sky.  N11 observe the SMC halo to be roughly symmetric in different azimuthal directions, and Fig~\ref{fig:break} shows that this property is suitably reproduced (i.e., the standard deviations of each model are comparable to the spread in observed data at each radius).

In order to facilitate comparison with N11, the models have been scaled to the observed data such that the density of model 1 matches the N11 datapoint at $r\approx2^{\circ}$.  The same scaling is applied to all three models.  We stress that our simulation neglects stellar evolution and cannot provide unique predictions for the density of red giants as observed by N11.  The scaling of the density profiles in Fig~\ref{fig:break} merely provides a visual comparison, and we justify this scaling as follows.  We cannot estimate the density of stars without arbitrarily assuming mass to light ratios, initial mass function, star formation history, etc., which is not possible in the present framework.  In addition, what is observed by N11 is only a fraction of the SMC stellar population, i.e., a portion of the red giant branch, which further separates the observed and simulated quantities.  Lastly, there are magnitude limitations that should be considered for the observed stellar densities.  As discussed in N11, previous studies (e.g. De Propris et al. 2010) failed to detect the extended SMC halo due to magnitude limitations, and it is feasible that future surveys utilizing deeper photometry than N11 may detect even more stars.  Despite these difficulties, the combined effect of these factors would likely change only the normalization of the radial density profile, so we choose to scale our models to the N11 data in Fig~\ref{fig:break} to facilitate comparison.

Judging from Fig~\ref{fig:break}, models 1 (cyan) and 2 (black) successfully populate the halo of the SMC at the large radial distances (up to $r\sim11^{\circ}$) observed by N11.  In addition, the models predict a ``break population" ($r>4^{\circ}$) having a much shallower slope for the radial density in comparison to the slope of the inner SMC population ($r<4^{\circ}$).  Moreover, the simulated slopes compare favorably to the observed values.  However, the densities for models 1 and 2 do not agree with the N11 data at all radii, and the predicted position of the break point is at a smaller radius than observed ($r\sim7^{\circ}$).  Nevertheless, models 1 and 2 are far more successful than model 3 (magenta), which fails to reproduce an extended halo population for the SMC.  Even though the slope of its radial profile in Fig~\ref{fig:break} abruptly changes at $r\sim4^{\circ}$ as in models 1 and 2, there are an insignificant number of particles stripped away in model 3.  Indeed, Fig~\ref{fig:new-zoom} shows that few particles are able to populate the outer regions of the SMC in model 3, and the distribution furthermore does not resemble the azimuthally symmetric halo discovered by N11.

Accordingly, we consider an extended spheroid like models 1 and 2 to better represent the tidal evolution of SMC.  Although the compact spheroid of model 3 is unable to reproduce the observations of Nidever et al. (2011), there is still the potential for unexplored intermediate cases 2.5 kpc $<R_{\rm sph, SMC}<$5.0 kpc.  In this sense, a compact spheroid is disfavored but not ruled out.

Before moving on, we compare our models to three more observational details of the SMC stellar halo: (i) N11 find that the population of red giants at radii $3^{\circ}<r<7.5^{\circ}$ has a center that is offset  by $\sim0.59^{\circ}$ from the center of the inner stellar distribution; (ii) N11 estimate the radial scale length of the SMC to be 0.8$^{\circ}$ (1.0$^{\circ}$) when using spherical (elliptical) model fits; (iii) The ellipticity of the SMC halo at $3^{\circ}<r<7.5^{\circ}$ is estimated by N11 to be $\epsilon \approx 0.1$, which is less flattened than the inner parts of the SMC ($\epsilon \approx 0.3$ at $r<3^{\circ}$, Harris \& Zaritsky 2006).

Comparing against the offset of centers (i), we find offsets of $\sim0.59^{\circ}$ for model 1, $\sim0.33^{\circ}$ for model 2, and $\sim0.20^{\circ}$ for model 3, where the offsets are calculated between the outer population ($3^{\circ}<r<7.5^{\circ}$) and inner population ($r<3^{\circ}$) in each model.  The cause of this offset can in general be attributed to the different responses to tidal perturbations between the inner and outer parts of the SMC, but perspective effects must be considered as stressed by N11.  To compare against the observed scale length (ii), we calculate the radial scale length in each of our spheroid models, which is roughly equal to $0.8 R_h$, where $R_h$ is the half-mass radius at the present day.  We find scale lengths of 0.6$^{\circ}$, 0.4$^{\circ}$, and 0.2$^{\circ}$ for models 1, 2, and 3, respectively.  Considering the ellipticity (iii), we find that the models exhibit the opposite behavior as observed, with greater spherical symmetry at small radii (where the gravitational potential is strongest) and more highly flattened distributions at large radii (where the potential is weaker and tidal disturbances are stronger).  From Fig~\ref{fig:new-zoom}, one can estimate that all models are roughly spherical $\epsilon \approx 0$ in the innermost region, and in the outer regions the ellipticity approaches $\epsilon \approx 0.3$ for models 1 and 2, and $\epsilon < 0.1$ for model 3.  In summary, we find points (i) and (ii) reinforce our preference for the extended spheroid models 1 and 2, whereas the discrepancies in ellipticity for point (iii) await the attention of future work.
}

\begin{figure} \centering
\includegraphics[scale=0.4]{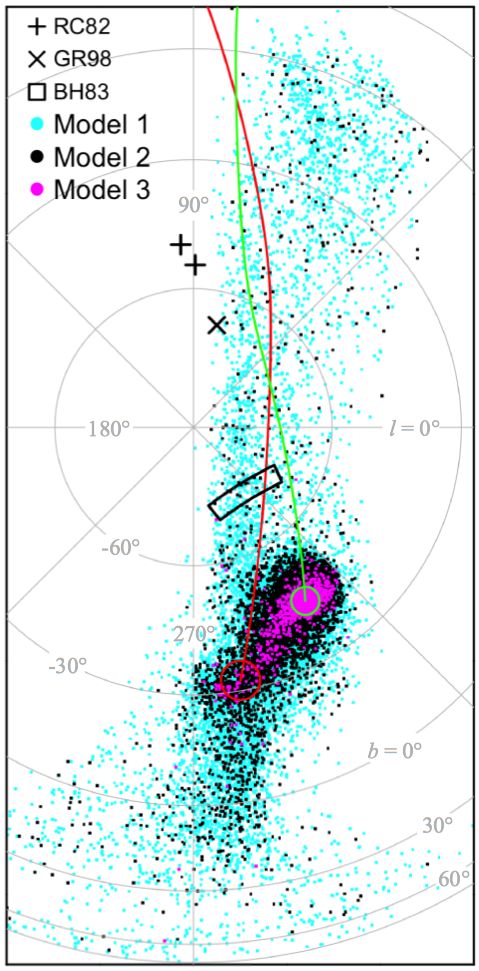}
\caption{The final distribution of spheroid particles projected onto the sky for models 1 (cyan), 2 (black), and 3 (magenta).  The location of survey fields that failed to find stars correlated with the Magellanic Stream are shown as plus signs (Recillas-Cruz 1982), a cross (Guhathakurta \& Reitzel 1998) and rectangle (Bruck \& Hawkins 1983).  The coordinates are galactic longitude $l$ (straight lines) and latitude $b$ (concentric circles) shown in a ZEA projection centered on the South Galactic Pole.  The orbit trails of the SMC (green) and LMC (red) are also shown, as are their current locations (circles).
\label{fig:spher-lb}}
 \end{figure}

\begin{figure*} \centering
\includegraphics[scale=0.5]{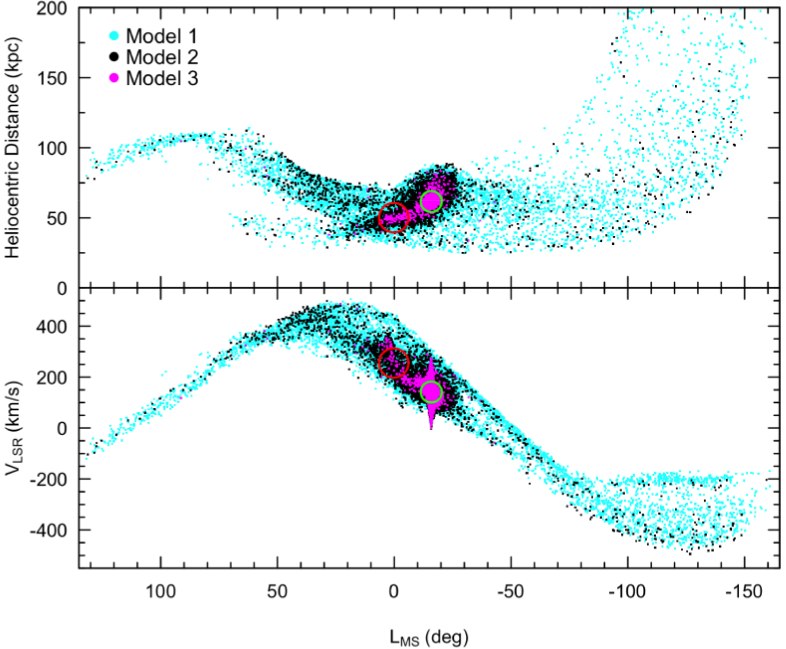}
\caption{The heliocentric distances (top panel) and $v_{\rm LSR}$ velocities (bottom panel) of the spheroid particles for models 1 (cyan), 2 (black), and 3 (magenta).  The horizontal axis is Magellanic longitude L$_{\rm MS}$, which runs parallel to the Magellanic Stream and is defined by Nidever et al. (2008).  In each panel, the leading material (i.e., analogue of the Leading Arm) is to the left of the SMC (green circle) and LMC (red circle), and the trailing material (i.e., analogue of the Magellanic Stream) is to the right.
\label{fig:spher-rhovlsr}}
 \end{figure*}

\subsection{On-sky Distribution and Distances} \label{sec:sph-sky}

In Fig~\ref{fig:spher-lb} we project the final distribution of the spheroid models onto the sky using a ZEA projection centered on the south galactic pole.  Comparing against Fig~\ref{fig:disk-coldens}, we see that the tip of the MS is offset from the stellar streams of models 1 and 2 (cyan and black, respectively).  This result is not unnatural because material removed from a rotating disk (e.g., the MS) may evolve differently to the material removed from a pressure-supported spheroid (e.g., the stellar stream). { Past observational attempts to locate stars along the MS have largely failed, such as Recillas-Cruz (1982; "plus signs" in Fig~\ref{fig:spher-lb}) and Guhathakurta \& Reitzel (1998; "cross" in Fig~\ref{fig:spher-lb}), who could not find stars at the tip of the MS.  Because the location of these survey fields are offset from the location of the streams of models 1 and 2 in Fig~\ref{fig:spher-lb}, we suggest that the Magellanic system may possess a stellar stream as long as it is displaced from the gaseous MS.  However, the existence of the stream can be challenged by surveys such as Bruck \& Hawkins (1983; rectangle in Fig~\ref{fig:spher-lb}), whose observations at the base of the MS overlap with the predicted streams of models 1 and 2.  Interestingly, the possible offset between the stripped gas and stars is echoed in the results of Bruck \& Hawkins (1983), because they observed a patch of stars at the base of the MS that did not correlate with the peak HI regions.}

If it exists, the stellar analogue of the MS may be difficult to observe.  As shown in the top panel of Fig~\ref{fig:spher-rhovlsr}, the distances to the stellar stream (L$_{\rm MS}<-20^{\circ}$, right hand side of Fig~\ref{fig:spher-rhovlsr}) are quite large, with many stars beyond 100 kpc at the tip of the stream.  { Even though the trailing stream of model 1 is quite dense ($1.2 \times 10^3~{\rm M}_{\odot}/{\rm deg}^2$, on average), the stream of model 2 is much more diffuse ($1.8 \times 10^2~{\rm M}_{\odot}/{\rm deg}^2$ on average).  Even though it is not clear what fraction of these simulated densities would correspond to observable stars, it is nevertheless certain that the stream of model 2 would be considerably more challenging to detect than that of model 1.}  The stellar analogue of the LA (L$_{\rm MS}>10^{\circ}$, left hand side of Fig~\ref{fig:spher-rhovlsr}) may also be difficult to observe due to confusion with the disk of the MW around $b\approx0$.

In contrast, the stripped stars in the Magellanic Bridge area should be relatively straightforward to observe.  Stars have indeed been located in the Bridge (e.g., Irwin et al. 1985), but they are young and blue, rather than old and red as would be expected from the spheroid population.  Indeed, Harris (2007) has suggested that the material stripped into the Bridge was almost purely gaseous, and any stars currently in the Bridge would have formed \emph{in situ}.  This may pose a problem in the current framework, as even our most compact SMC spheroid (model 3, $R_{\rm sph, SMC}=2.5$ kpc) is easily stripped into the Bridge region, { where the average density is $3.6 \times 10^3~{\rm M}_{\odot}/{\rm deg}^2$.  The corresponding Bridge densities for models 1 and 2 are $5.7 \times 10^4~{\rm M}_{\odot}/{\rm deg}^2$ and $2.6 \times 10^4~{\rm M}_{\odot}/{\rm deg}^2$, respectively.} We comment further on this topic in the Discussion Sec~\ref{sec:dx-stars}.

\subsection{Kinematics and SMC Velocity Profile}
The bottom panel of Fig~\ref{fig:spher-rhovlsr} gives the radial velocity profile of the stripped spheroid material.  The profile for the trailing stream of models 1 and 2 (L$_{\rm MS}<-20^{\circ}$, right hand side of Fig~\ref{fig:spher-rhovlsr}) is largely coincident with that of the stripped disk particles (Fig~\ref{fig:disk-rhovlsr}, lower panel), and follows the kinematic trends of both the MS and Horn.  { The trailing streams of models 1 and 2 also exhibit an apparent ``pile-up" of particles with $v_{\rm LSR} \approx -200$ km~s$^{-1}$ between L$_{\rm MS}=-100^{\circ}$ and $-150^{\circ}$, which mirrors the velocity inflection at the tip of the gaseous MS (Nidever et al. 2010), but it is important to note that these particles are distributed over a large range of distances (Fig~\ref{fig:spher-rhovlsr}, upper panel).}

The velocity profile for the leading stream (L$_{\rm MS}>10^{\circ}$, left hand side of Fig~\ref{fig:spher-rhovlsr}) is similar to that of the simulated LA (Fig~\ref{fig:disk-rhovlsr}, lower panel), but it is considerably more extended to lower velocities at L$_{\rm MS}>70^{\circ}$.  The particles for model 3 are stripped only into the Bridge region, as is clear from Fig~\ref{fig:spher-rhovlsr}.  We remind the reader that the tidal evolution of the \emph{disk} failed to explain the small rotational amplitude of the intermediate-age stars of the SMC (i.e., as measured by HZ06; see Section ~\ref{sec:disk-smcv}).  To reveal the velocity structure of stripped SMC stars (e.g., Fig~\ref{fig:spher-rhovlsr}) future models must therefore represent the SMC as a ``spheroid plus disk" as in the present work, departing from the traditional ``pure disk" models of the past (e.g., Connors et al. 2006; Besla et al. 2010).

The SMC velocity profile for each of the spheroid models is given in Fig~\ref{fig:disk-vel}.  Models 1, 2, and 3 (panels b, c, d, respectively) all exhibit a velocity gradient of $\sim$10 km s$^{-1}$ deg$^{-1}$ across the face of the SMC.  This corresponds nicely to the observations of Harris \& Zaritsky (2006), who report a similar gradient of 8.3 km s$^{-1}$ deg$^{-1}$.  In their analysis, HZ06 determine that the maximum possible rotation velocity that could be derived from this velocity gradient is significantly less than the velocity dispersion.  For this reason, HZ06 conclude that the the stellar component of the SMC is a dispersion supported spheroid.  All three of our models are consistent with this analysis, but one may still ask \emph{why} the velocity gradient is imprinted into the SMC.  { Harris \& Zaritsky (2006) consider that that the observed velocity gradient may be caused by a projection effect of the SMC space velocity along the line of sight.  That is, because the SMC occupies a large area on the sky, the projection of radial velocity at different regions of the SMC will necessarily vary (e.g., van der Marel et al. 2002).

In contrast, we consider that the velocity gradients in our simulations are due to tidal disturbances.}  As we have shown previously, much material is stripped away from the SMC during the close passages of the LMC at $t=-1.97$ Gyr and $t=-0.26$ Gyr.  What we emphasize here is that these close encounters also strongly affect the particles that remain within the SMC.  In particular, each close passage of the LMC effectively reorganizes the spheroid particles along a velocity gradient.  If the particles are pulled beyond the tidal radius of the SMC, as in the case of models 1 and 2 for the $t=-1.97$ Gyr passage, then the velocity gradient determines how the leading and trailing streams separate from the SMC body.  Even though the particles of model 3 remain bound to the SMC during this interaction, the entire spheroid is nevertheless imprinted with a velocity gradient.  As shown at the present day in Fig~\ref{fig:disk-vel} (i.e., at $t=0$ Gyr), the spheroids exhibit velocity gradients that are the signature of the most recent LMC passage at $t=-0.26$ Gyr.

\subsection{Accretion onto the LMC} \label{sec:sph-inlmc}

\begin{figure}[t] \centering
\includegraphics[trim=55 40 0 400, scale=0.5]{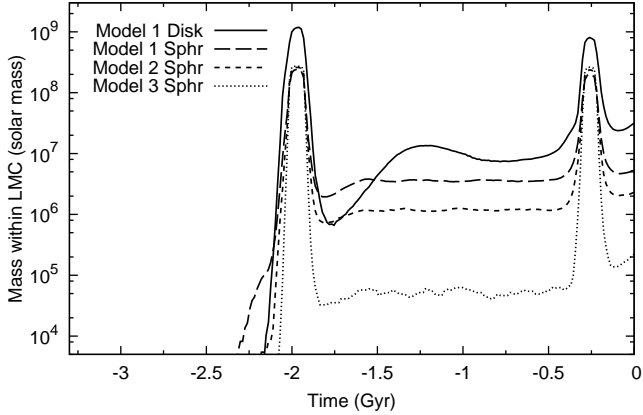}
\caption{The total mass falling within 7.5 kpc of the LMC as a function of time, shown for the SMC disk of model 1 (solid line), and the SMC spheroid of models 1 (long dash), 2 (dash), and 3 (dotted).  The two peaks correspond to the two close passages of the LMC and SMC.
\label{fig:inlmc}}
 \end{figure}

Fig~\ref{fig:inlmc} shows the total mass falling into LMC ($r<7.5$ kpc) as function of time for the SMC disk (model 1) and the SMC spheroid (models 1, 2, and 3).  { We stress that Fig~\ref{fig:inlmc} merely gives the particles passing through the LMC at each time step, whereas the fate of each particle (i.e. bound or unbound to the LMC) depends on its energy.}  The mutual close encounters of the MCs at $t=-1.97$ Gyr and $t=-0.26$ Gyr bring their centers of mass to within 6.0 kpc and 6.6 kpc, respectively, which explains the large temporary peaks in Fig~\ref{fig:inlmc}.  However, a portion of the SMC particles are truly transferred to the LMC following each epoch of interaction.  Up to the present day ($t=0$ Gyr), the total mass transfered to the LMC equals $\sim3.1\times10^7$ M$_{\odot}$ from the SMC disk, $\sim5\times10^6$ M$_{\odot}$ from the extended spheroid (model 1), $\sim2\times10^6$ M$_{\odot}$ from the intermediate spheroid (model 2), and $\sim2\times10^5$ M$_{\odot}$ from the compact spheroid (model 3).  For each component, approximately half of the mass was captured by the LMC following the first interaction at $t=-1.97$ Gyr and the other half following the second interaction at $t=-0.26$.  As can be seen in Fig~\ref{fig:inlmc}, the transfer of mass from the disk exhibits a sharp rise from $\sim10^6$ M$_{\odot}$ at $t=-1.75$ Gyr to $\sim1.3\times10^7$ M$_{\odot}$ at $t=-1.25$ Gyr, in contrast to the comparatively flat distributions of mass transfer from the spheroids.  This can be attributed to the strong rotation of the disk particles as they fall gradually into the LMC with considerable angular momentum.  It also explains the slight decrease of mass transfer from the disk between $t=-1.25$ Gyr and $t=-0.8$ Gyr because some particles that initially fall within 7.5 kpc of the LMC subsequently exceed this radius by virtue of their large angular momentum.

\begin{figure} \centering
\includegraphics[trim=20 0 0 0, scale=0.3]{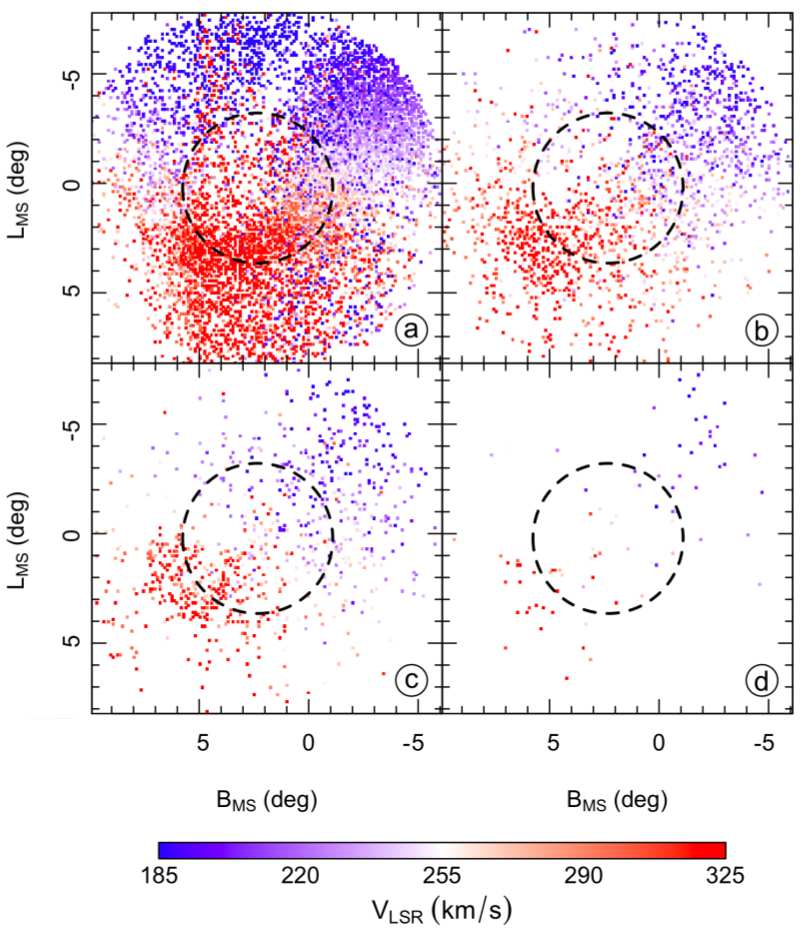}
\caption{The $v_{\rm LSR}$ velocity field of particles accreted to within 7.5 kpc of the LMC.  The particles originate from (a) the SMC disk of model 1, and (b-d) the SMC spheroid of models 1-3, respectively.  Each panel is centered on the LMC, and the dashed circle outlines a radius of 3 kpc.  The on-sky coordinates are Magellanic longitude L$_{\rm MS}$ and latitude B$_{\rm MS}$ as defined by Nidever et al. (2008).  The velocity range has a lower bound of 185 km s$^{-1}$ (blue), an upper bound of 325 km s$^{-1}$ (red), and is centered on the $v_{\rm LSR}$ velocity of the LMC (white).
\label{fig:inlmcv}}
 \end{figure}

Recently, Olsen et al. (2011) have discovered a population of AGB stars within the LMC that possess peculiar kinematics and metallicities.  They suggest that these stars may have originated in the SMC and were accreted at some point during the recent dynamical interaction with the LMC.  Our models provide a concrete realization of this scenario.  In particular, because AGB stars would be represented by the old type stellar population of the SMC spheroid, the transfer of spheroid particles to the LMC reproduces the main result of Olsen et al. (2011).  Fig~\ref{fig:inlmcv} gives the $v_{\rm LSR}$ velocity profile of the particles captured by the LMC (i.e., $r<7.5$ kpc at $t=0$ Gyr) from the SMC disk (panel a), and the spheroids of models 1, 2, and 3 (panels b, c, and d, respectively).  The properties of the accreted spheroid particles are broadly consistent with the results of Olsen et al. (2011), { namely because the kinematics are distinct from that of the LMC disk.  However, the exact kinematical signature of the Olsen et al. (2011) population, i.e., \emph{counter-rotation} with respect to the LMC disk, is not reproduced.}  Considering that the population of stars that Olsen et al. (2011) attribute to the SMC comprise a small but significant percentage $\gtrsim5\%$ of their observed sample, the mass transfer scenario of model 3 (Fig~\ref{fig:inlmcv}, panel d) may be insufficient to explain the observation.  This would seem to argue against a compact spheroid for the SMC.  In Sec~\ref{sec:dx-stars} we consider further implications of the Olsen et al. (2011) observation in the context of other results in the literature.

\section{Discussion} \label{sec:dx}

\subsection{The Possibility of Tidally Stripped SMC Stars} \label{sec:dx-stars}

{ Numerous surveys have shown that stars do not exist in large numbers along the MS and furthermore that they do not correlate with peak HI regions (e.g., Recillas-Cruz 1982; Bruck \& Hawkins 1983).  Additionally, Guhathakurta \& Reitzel (1998) use the null observation of stars in their $5' \times 7'$ field to derive an upper bound on the ratio of stellar to gaseous mass $M_{\star}/M_{\rm HI}\approx$0.1 at the tip of the MS.  Is the present model consistent with these constraints?  We must be careful to consider two distinct populations when discussing the tidal evolution of SMC stars: (i) the stellar component of the disk, and (ii) the spheroidal component of old-type stars.  As seen in Fig~\ref{fig:spher-lb}, stars stripped from the spheroid do not coincide with the MS tip, nor is there a significant overlap with the probes of Recillas-Cruz (1982) or Guhathakurta \& Reitzel (1998).  In addition, there would be no reason for stars stripped from the spheroid to correlate with the high-density regions of the MS, which is consistent with the above observations (Sec~\ref{sec:sph-sky}).  Stars stripped from the disk, however, would be interspersed among the gas clouds of the MS and would likely coincide with peak HI regions, which poses a problem if the stripped stars are too numerous.  In other words, stars stripped from the SMC spheroid may be permitted by current observational constraints, but those sourced from the SMC disk face stronger restrictions.}

\begin{figure}[t] \centering
\includegraphics[trim=55 40 0 420, scale=0.5]{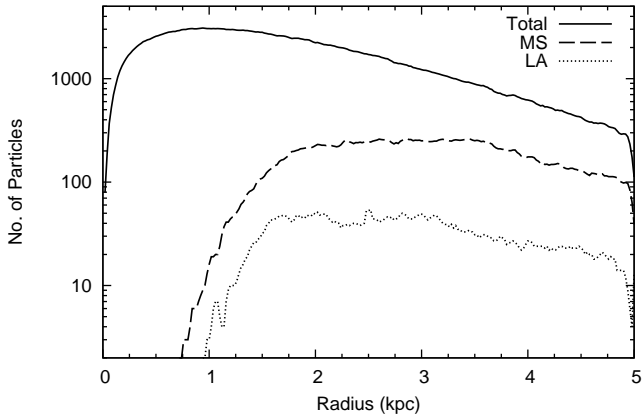}
\caption{Number of particles plotted as a function of their original radius within the SMC disk.  The solid line corresponds to all particles, whereas the dashed and dotted lines refer to the particles that constitute the Magellanic Stream (MS) and Leading Arm (LA), respectively.  The MS is selected as all particles satisfying L$_{\rm MS}<-25$ that are not in the Horn, and the LA is selected as all particles satisfying L$_{\rm MS}>5$ and having distances of greater than 50 kpc.
\label{fig:origin}}
 \end{figure}

To address this concern, previous models have assumed that the stellar component of the SMC disk was confined to the central regions where tidal stripping is inefficient (Yoshizawa \& Noguchi 2003; Besla et al. 2010).  In the present work, we naively associated the collisionless disk particles of our model with gas (Sec~\ref{sec:mod-nbody-smc}), but here we consider the possibility that some portion of the disk, particularly the central regions, may pertain to stellar material.  In Fig~\ref{fig:origin} we plot the number of disk particles as a function of their radius within the initial SMC disk (solid line), and we also show the fraction that eventually constitute the MS (dashed) and the LA (dotted).  The particles are selected to be in the MS or LA based on the criteria given in the caption of Fig~\ref{fig:origin}.  Most of the stripped particles originate in the outer half of the SMC disk, with the average radius being 3.04 kpc and 3.03 kpc for the MS and LA, respectively.   If we assume that the disk mass within $r=1.75$ kpc (2.0 kpc) is completely stellar and beyond this radius is strictly gaseous, then the stellar to gas mass ratio is 0.09 (0.19), which is broadly consistent with the upper limit of $\sim$0.1 measured by Guhathakurta \& Reitzel (1998) at the MS tip.  This would imply that the initial SMC gas disk ($r=$5.0 kpc) had an extent which was 2.5 to 2.85 times that of the stellar component in the disk, which is not uncommon for dwarf galaxies (Swaters et al. 2002).

{ Even though there is no direct evidence for a stellar counterpart to the MS, numerous studies indicate that the outer halos of the LMC and SMC are much more expansive than previously appreciated (e.g., Kunkel et al. 1997b; Noel \& Gallart 2007; Saha et al. 2010; De Propris et al. 2010), extending more than 20$^{\circ}$ from the LMC (Mu\~noz et al. 2006; Majewski et al. 2009) and more than 10$^{\circ}$ from the SMC (Nidever et al. 2011).  It would seem natural for these extended halos to be accompanied by diffuse tidal streams of stars, and there are hints that such structures may indeed exist.  For example, stars observed in the foreground of Carina may be the first evidence of tidal stripping from the SMC and/or LMC stellar halo (Mu\~noz et al. 2006).  In addition, the radial density profiles of the LMC stellar periphery (Majewski et al. 2009) and that of the SMC (Nidever et al. 2011, hereafter N11) both exhibit a sharp ``break" beyond which the slope of the density profile changes.  Such a break is present in many other tidally evolved systems and may indicate the presence of tidal tails, as in the Sagittarius dwarf galaxy (Majewski et al. 2003), the Carina dwarf spheroidal galaxy (Mu\~noz et al. 2006), and numerous globular clusters (Chun et al. 2010).  However, the observation of a break population cannot by itself establish the existence of tidal structures, because it may also be consistent with a bound stellar halo (e.g., Mu\~noz et al. 2008).}

{ In the present study, we are better able to reproduce the SMC break population (e.g., Fig~\ref{fig:break}) when we adopt an extended stellar spheroid ($R_{\rm sph, SMC} \gtrsim 5$ kpc) rather than an initially compact one ($R_{\rm sph, SMC} = 2.5$ kpc).  In turn, the model predicts that diffuse tidal features are stripped away from the extended spheroid during its tidal evolution, forming stellar structures analogous to the gaseous MS, LA, and Bridge.  In other words, we have used the N11 observations of the SMC stellar periphery to support the hypothesis that stellar tidal structures exist in the Magellanic system.  Future wide-field observational surveys such as the Southern Sky Survey by the SkyMapper telescope (Keller et al. 2007) would be ideal for detecting the stellar counterparts to the MS and LA, if they indeed exist, and establishing a connection to the outer halo of the SMC.}

{ The present prediction that stars should be stripped into the Bridge region from all spheroid models (Sec~\ref{sec:sph-sky}) is in conflict with Harris (2007), who observed the density of old-type stars in the Bridge to be quite low.  Young blue stars have indeed been observed in the Bridge, but they were not tidally stripped from the SMC but rather were formed in situ following the epoch of tidal stripping (Irwin et al. 1985).  It would appear that a much more compact spheroid ($R_{\rm sph, SMC} \lesssim 2.5$ kpc) would be needed to avoid stripping into the Bridge, but such a spheroid would unlikely be able to reproduce the extended halo structure observed by N11.  Furthermore, the recent contention of Olsen et al. (2011) that the LMC has accreted a population of AGB stars from the SMC also creates tension with Harris (2007), because we find that tidal stripping through the Bridge region provides a mechanism of such a mass transfer (see sections~\ref{sec:sph-sky} and~\ref{sec:sph-inlmc}).  Another curiosity follows from the large line-of-sight depth of the SMC, which has been explained in the present model as a projection of the tidally extended Counter-Bridge (Sec~\ref{sec:disk-bcb}).  However, this explanation relies on \emph{stars} being tidally stripped into the Counter-Bridge, which would logically imply that stars should be stripped into the Bridge as well.  Regardless of these details, it is clear that many recent studies (including the present work) have placed a renewed interest in the tidal history of the SMC stellar population.}

\subsection{Hydrodynamical Interactions with the Milky Way's Hot Halo} \label{sec:dx-la}

The present work shows that gravitational interactions alone can explain the morphology and kinematics of the MS and Bridge, but the LA on the other hand is not well explained in the current framework.  Here we consider the second-order effects of including hydrodynamics and its potential for reproducing the LA.  The major hydrodynamical interactions to be considered are encompassed by the collision of neutral gas clumps (e.g., within the MS and LA) with the tenuous hot halo of the Milky Way.  The drag induced by this interaction is termed ``ram pressure" and is proportional to $\rho v^2$ where $\rho$ is the density of the hot halo at the location of the HI gas clump, and $v$ is the relative velocity between the hot halo and the clump (e.g., Gunn \& Gott 1972; Westmeier et al. 2011).  As a simple approximation, the density $\rho$ can be considered a function of galactocentric radius $r$, with reasonable choices scaling as $\rho \sim r^{-2}$ (isothermal) or $\rho \sim r^{-3}$ (NFW), and the velocity $v$ can be taken as the galactocentric velocity if we assume the hot halo is non-rotating.  These simple assumptions will allow us to calculate the relative effect of ram pressure on the MS and LA throughout their orbit about the MW.

\begin{figure} \centering
\includegraphics[trim=40 40 0 250, scale=0.48]{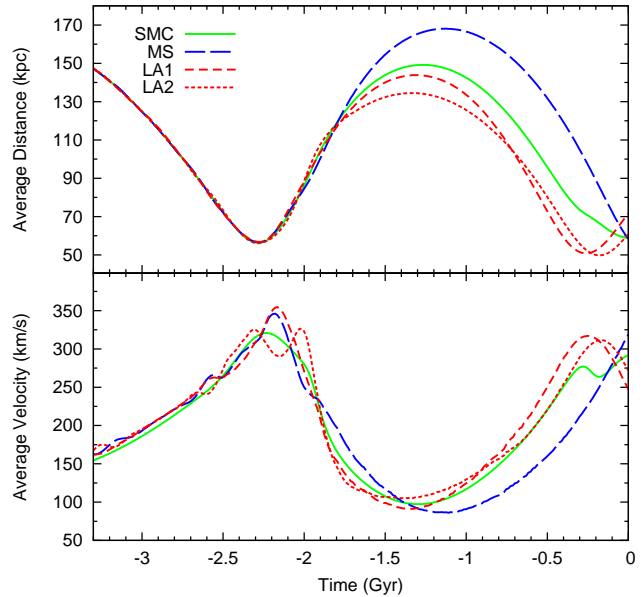}
\caption{The time evolution of the average radius (top panel) and velocity (bottom panel) of the SMC (green solid line) and its various stripped components, including the Magellanic Stream (MS; blue long dashed line) and the two branches of the Leading Arm (LA1 and LA2; red dashed and red dotted line, respectively).  The MS and LA are selected as in Fig~\ref{fig:origin}, with an additional division of the LA according to heliocentric distances and $v_{\rm LSR}$ velocities.  LA1 is the branch having larger distances and slightly larger radial velocities; it is colored yellow-orange in Fig~\ref{fig:disk-rhovlsr} and originates in tidal arm A.  LA2 has smaller distances and lower radial velocities; it is colored black in Fig~\ref{fig:disk-rhovlsr} and originates in tidal arm B.
\label{fig:rvt}}
 \end{figure}

Fig~\ref{fig:rvt} gives the galactocentric distance (top panel) and average velocity (bottom panel) throughout the entire orbital evolution of the present model.  The MS and LA are selected as in Fig~\ref{fig:origin}, but the LA is split into two branches (LA1 and LA2) according to their distinct evolutionary tracks.  As can be seen, the orbits of the LA branches are generally faster and at smaller galactic radii than that of the MS, and the corresponding effect of ram pressure will accordingly be larger.  For example, at $t=-0.3$ Gyr, one can calculate that the relative effect of ram pressure on the LA should be at least an order of magnitude larger than for the MS.  Guided by this simple calculation, we consider hydrodynamical effects to be relatively minor for the MS but very important for determining the morphology and kinematics of the LA.  This contention will be explored with future, more sophisticated N-body models.  In our previous work (DB11b) we argued that the distances and velocities of the LA can decrease in response to a simplistic prescription for ram pressure, and that such effects would produce an improved agreement with observations (e.g., McClure-Griffiths et al. 2008).  Under these expectations, strong ram pressure would likely cause the LA to sink to lower radii subsequent to $t=-0.3$ Gyr in the present model.  However, because we adopt collisionless dynamics in the present work, the LA is instead predicted to rise to much larger radii subsequent to $t=-0.3$ Gyr (Fig~\ref{fig:rvt}).  For this reason, we cannot consider the present collisionless model to adequately capture the evolution of the LA.  This explains, at least partially, why the predicted on-sky location in Fig~\ref{fig:disk-coldens} is quite different to that observed.

Hydrodynamical effects at the interface of the hot halo are also responsible for the shredded, ionized filaments of the MS (Stanimirovi\'c 2008; Westmeier \& Koribalski 2008), the disjointed arrangement of LA clouds and their head-tail structure (Bruns et al. 2005), and the anomalous H$\alpha$ emission within the MS (Putman et al. 2003b; Bland-Hawthorn et al. 2007).  All of these interactions lead to a decrease in the overall HI mass by transforming the neutral gas to an ionized component.  The observed HI masses of the MS and LA are $1.2\times10^8$ M$_{\odot}$ and $0.3\times10^8$ M$_{\odot}$ (Bruns et al. 2005), respectively, and the corresponding predicted masses of the present model are $1.2\times10^8$ M$_{\odot}$ and $0.2\times10^8$ M$_{\odot}$, where we have used the definitions of Fig~\ref{fig:origin} to select the MS and LA.  Note that Putman et al. (2003a) quote a larger MS mass of $1.9\times10^8$ M$_{\odot}$ due to a larger region of selection for the MS.  Even though the predicted masses compare well with the measured values, the inclusion of hydrodynamical interactions would have the unwanted effect of reducing the predicted HI masses.  This discrepancy could be combated by increasing the initial SMC mass, but this would alter the orbital history and may not be favorable for the formation of the MS.  Another possible solution is to change the mass ratio of the gas disk while keeping the total SMC mass fixed.  We leave this nontrivial issue for future work.

\subsection{Mass Transfer from the SMC to the LMC} \label{sec:dx-inlmc}

The present models have clearly shown that gas and stars in the SMC can be transferred to the LMC via tidal interaction over the last $\sim$2 Gyr.  In particular, two epochs of significant mass transfer are predicted, one following the first strong interaction between the MCs at $t=-1.97$ Gyr, and the other following the second interaction at $t=-0.26$ Gyr.  Here we consider how this predicted scenario compares with observational evidence.

\subsubsection{Gas Transfer to the LMC}  \label{sec:dx-inlmc-gas}

Through collisionless dynamics alone, our model predicts that the transfer of \emph{gas} to the LMC (i.e., transfer from the SMC \emph{disk}) totaled $\sim1.3\times10^7$ M$_{\odot}$ by $t=-1.25$ Gyr and $\sim3.1\times10^7$ M$_{\odot}$ by the present day.  However, these figures would increase, perhaps dramatically, once hydrodynamics is considered, because collisional forces would precipitate the infall of much material that otherwise passes straight through the LMC at $t\approx-2.0$ Gyr and $t\approx-0.25$ Gyr (i.e., the two temporary peaks in Fig~\ref{fig:inlmc}).  Owing to the fact that the gas from the SMC is relatively metal-poor, the accretion of this gas onto the LMC would provide fuel for the formation of low-metallicity stars (Bekki \& Chiba 2007).  The prediction that such a process has occurred is supported by two separate pieces of observational evidence: (i) the chemical enrichment history of the LMC, and (ii) the unusually low abundance of nitrogen in young stellar populations of the LMC.

As described in Sec~\ref{sec:mod-orb}, we used the correlated star formation histories of the LMC and SMC (Harris \& Zaritsky 2009) to constrain the timing of their close tidal interactions for the past $\sim2$ Gyr.  But how did the metallicity of the LMC evolve during this time?  { The chemical enrichment history of the LMC is provided by numerous authors (e.g., Figure 20 of Harris \& Zaritsky (2009); Fig 1 of van Loon et al. 2005; and the ongoing VMC survey, Rubele et al. 2012), which shows that the average metallicity of the LMC has generally increased over its $\sim10^{10}$ Gyr lifetime but has \emph{decreased} during the most recent $\sim2$ Gyr.}  Considering Fig~20 of Harris \& Zaritsky (2009), the metallicity has decreased during two epochs: between $\sim2$ to $\sim1$ Gyr ago, and between $\sim500$ Myr ago to the present day.  These curious trends could possibly be explained by widespread star formation from relatively metal-poor gas ([Fe/H]$<-1$), and the accretion of such gas onto the LMC is indeed predicted by the present model.  This gas would likely admix into the LMC and effectively dilute the interstellar medium from which new stars would form.  Notice also that the periods of decreasing metallicity in the LMC are remarkably well aligned with the two epochs of correlated star formation ($\sim2$ Gyr ago and $\sim500$ Myr ago; Harris \& Zaritsky 2009).  In fact, one can suggest that the decrease in metallicity appears to be a \emph{response} to the bursts of star formation.  Following the predictions of the present model, the two strong tidal encounters between the MCs (at $t=-1.97$ and $t=-0.26$ Gyr) would have caused increased star formation in each galaxy, which explains the correlated bursts of star formation.  And because the metal-poor gas from the SMC is transfered to the LMC in response to each tidal encounter, the stars in the LMC would subsequently form with relatively lower metallicities, which explains the decreasing trends in the chemical enrichment history.  However, we should point out that the dominant mechanism for triggering the star formation in the LMC is not at all clear; it was likely a competition between the tidal interaction with the SMC and the gas infall from the SMC.

Recent observations have confirmed that the chemical abundances of star-forming HII regions and very young stellar populations in the LMC have [N/H] ratios that are a factor
of $6-7$ lower than the solar value (e.g., Korn et al. 2002; Hill 2004; van Loon et al. 2010).  On the other hand, other elements such as O and Ne are underabundant by a factor of only $\sim 2$ in these systems (Hill 2004).  In other words, the young stellar populations of the LMC exhibit curiously low abundances of nitrogen.  To explain this anomaly, Bekki \& Tsujimoto (2010) have invoked recent gas accretion onto the LMC, and they find that the most likely origin of this gas was from the SMC rather than, for example, high-velocity clouds (HVCs) in the Local Group.  { In particular, this gas would have originated in the \emph{outer} parts of the SMC where metallicities were low owing to a negative metallicity gradient (Bekki \& Tsujimoto 2010).} The present model provides a concrete realization of this scenario as long as we assume that the stripped gas is deficient in nitrogen.  This assumption can be justified by the following argument.  Much of the gas that is accreted onto the LMC in our model has a similar origin as the MS, namely, as the tidally stripped material in the outer regions of the SMC disk, and it stands to reason that their chemical abundances should also be similar.  Given that Fox et al. (2010) have observed very low nitrogen abundances in the MS, it is not unreasonable to assume that the gas transfer to the LMC was similarly nitrogen-poor.  Under this assumption, the present model strengthens the conclusions of Bekki \& Tsujimoto (2010), although it is unclear what \emph{amount} of nitrogen-poor gas must be transfered to explain the properties of the young stellar populations of the LMC.  We leave this quantitative investigation to our future work.

\subsubsection{Kinematically Peculiar Stars in the LMC} \label{sec:dx-inlmc-stars}

Olsen et al. (2011) have recently investigated the distribution and kinematics of AGB stars in the LMC and found that about 5\% of the stars in their sample have line-of-sight velocities that oppose the sense of rotation of the LMC disk.  It is unlikely that these stars formed in the LMC, because in addition to having peculiar kinematics, they have much lower metallicities ([Fe/H]$=-1.25 \pm 0.13$) than the field stars of the LMC ([Fe/H]$= -0.56 \pm 0.02$).  Olsen et al. (2011) conclude that the most likely origin for these stars was the SMC, principally because De Propris et al. (2010) measured a similar metallicity of [Fe/H]$= -1.35 \pm 0.10$ for the red giants in the periphery of the SMC.  Two HI arms within the LMC are kinematically linked to the anomalous AGB population, and though these gaseous structures were previously identified as outflows (Staveley-Smith et al. 2003), they are { hypothesized} by Olsen et al. (2011) to be infalling material.  The present model reinforces this general scenario and illustrates a dynamical process whereby old-type stars in the SMC spheroid are tidally stripped and engulfed by the LMC.  The model also predicts that the stripped components of the SMC (both disk and spheroid) can exhibit rotation in the outer part of the LMC, possibly explaining the observed out-of-plane stellar polar ring (Kunkel et al. 1997a) and the counter-rotating stellar component in the LMC (Subramanian \&  Subramaniam 2009).

The transfer of stars from the SMC to the LMC occurs during two different epochs ($t\approx-2.0$ Gyr and $t\approx-0.25$ Gyr) following the close encounters between the MCs.  The second epoch corresponds to the formation of the Magellanic Bridge, and the stripped stars from the SMC spheroid should indeed coincide spatially with the gas in the Bridge.  Even the most compact spheroid model that we adopt is easily stripped into the Bridge region (Sec~\ref{sec:sph-sky}), but this prediction may conflict with previous failed attempts to detect old-type stars within the Bridge (e.g., Harris 2007).  We should point out that tidally stripping stars through the Bridge region provides an obvious mechanism of mass transfer to the LMC, but it is not the only mechanism in our model, since considerable mass is also transfered during the first interaction between the MCs at $t\approx-2.0$ Gyr.

In the present model, the total stellar mass that the LMC accretes is at most $\sim 5\times10^6 M_{\odot}$, which comes from adopting an extended spheroid (Model 1, $R_{\rm sph}=7.5$ kpc).  Though Olsen et al. (2011) claim an SMC origin for 5\% of their sample, it is not immediately clear if the total mass of this population within the LMC should be estimated as 5\% of the current LMC stellar mass $\sim3 \times 10^9 M_{\odot}$ (van der Marel et al. 2002).  If we do make this simple assumption, then the total mass of SMC stars transfered to the LMC should be  $\sim 1.5 \times 10^8 M_{\odot}$.  This figure is a factor of $\sim30$ larger than predicted, and is in fact comparable to the total initial mass that we assume for the SMC spheroid, $M_{\rm sph}=2.7 \times 10^8 M_{\odot}$.  This may imply that the observational estimate of the mass fraction of accreted SMC stars is overestimated, or it may suggest that a different accretion/merger event for the LMC (e.g., minor merging of small satellite galaxies) is needed to explain the origin of the kinematically peculiar AGB stars.  A final possibility is that the SMC may have had a significantly larger total mass before it began to interact with the LMC and Milky Way, which may be reasonable given that a recent study of the SMC's rotation curve proposed a larger mass of more than $6.5 \times 10^9 M_{\odot}$ (Bekki \& Stanimirovi\'c 2009).

We accordingly suggest that an estimate of the total mass of the accreted SMC populations may help to constrain the original SMC mass prior to its interactions with the LMC.  Although the only observations of possible accreted SMC populations is the AGB dataset of Olsen et al. (2011), ongoing photometric and spectroscopic studies of planetary nebulae in the LMC will enable the determination of other accreted populations from the SMC.  Moreover, these future studies will allow us to discuss how the total masses of the accreted SMC populations might depend on stellar ages and abundances.  In addition, the kinematically peculiar HI gas in the LMC can be used to give further constraints on the mass transfer process between the LMC and the SMC, as suggested by Olsen et al. (2011).  In the future, we plan to run chemodynamical simulations to discuss how the gas and stars of the SMC can be transferred to the LMC and consequently influence its structure and dynamics.

\subsection{Diagnosis for recent MS formation models: Success and Failure} \label{sec:dx-compare}

The formation of the MS has been described by numerous dynamical models in recent years, and here we will point out (i) the key differences among their methodologies, and (ii) their relative success and failure to explain the observed properties of the MCs, MS, Bridge, and LA.  We divide the models into four groups: the traditional tidal models (e.g., C06), the ram pressure stripping model of Mastropietro et al. (2005, hereafter M05), the ``first passage" model of Besla et al. (2010, hereafter B10), and the present tidal model.  Although there are a number of MS models based on test particle simulations (e.g., Ruzicka et al. 2010; DB11a), we focus here on the N-body and hydrodynamical simulations.  Our comparison is summarized graphically in Table~\ref{tab:compare}.  It should be stressed that we are simply attempting to highlight the relative strengths of the models rather than trying to discern which one is the ``best."  After all, each model will certainly be improved in the future as more realistic physical prescriptions are added and higher resolution simulations become accessible.

Reproducing the orbital constraints of the MCs is one of the fundamentally important tasks for any model of the Magellanic system.  The present work adopts the values given in Table~\ref{tab:param1} for the on-sky positions, distances, line-of-sight velocities, and proper motions of the MCs.  Of these parameters, only the proper motions are poorly constrained, although recent measurements have suggested that the low-velocity LMC orbits adopted by C06 and M05 may not be realistic.  The high-velcoity LMC orbit adopted by B10 is justified by the proper motion measurements of Kallivayalil et al. (2006a), and the LMC orbit of the present model is supported by Vieira et al. (2010).   While all models reproduce the LMC orbit (barring the difference in proper motions), the same is not true for the SMC.  Perhaps for simplicity, M05 neglect the SMC entirely in their model, focusing only on the interaction between the LMC and the Milky Way.  Because the present model shares the same methodology as C06, namely the backward orbit integration scheme, the SMC orbit can be easily reproduced by suitable choice of parameters.  The task is nontrivial, however, for the numerical method adopted by B10, and consequently their orbit for the SMC has the incorrect on-sky position, line-of-sight velocity, and proper motion (see Besla et al. 2012).

The underlying reason that B10 fail to reproduce the correct SMC orbit is their adoption of self-consistent interactions between the MCs, i.e., they represent both the LMC and SMC as ensembles of N-body particles.  This choice certainly makes the B10 model more sophisticated than either C06 or the present work, which instead adopt a fixed potential for the LMC, but it also implicitly introduces dynamical friction between the MCs, rendering pre-calculation of the orbits impossible.  To clarify: in C06 and in the present work, present-day orbital constraints can be \emph{chosen} as initial conditions for orbit integrations, and the N-body evolution is guaranteed to follow these pre-calculated orbits; in self-consistent models, however, the N-body evolution will not obey a pre-calculated orbit.  Instead, one must run many full-resolution models to test which initial conditions, if any, reproduce the present-day orbital constraints.  Add to this the requirement that the orbit deliver a good tidal stripping model for the MS, and it is understandable why B10 were unable to satisfy all constraints.

\begin{deluxetable}{ccccc}[t]
\footnotesize
\tablecaption{Comparison of recent MS models
\label{tab:compare}}
\tablewidth{0pt}
\tablehead{
\colhead{  Properties/models  \tablenotemark{0} } &
\colhead{  C06  \tablenotemark{1} } &
\colhead{  M05 \tablenotemark{1} } &
\colhead{  B10  \tablenotemark{1} }  &
\colhead{  This work  }  }
\startdata

LMC orbital constraints    &  $(\surd)$  &  $(\surd)$  &  $\surd$  &  $\surd$  \\
SMC orbital constraints    &  $(\surd)$  &  $-$  &  $-$  &  $\surd$  \\
Self-consistent interactions &  $-$  & $\surd$  &  $(\surd)$  &  $-$  \\
Realistic MW potential      &  $-$  & $\surd$  &  $\surd$ & $\surd$ \\
MW hot halo interaction   &  $-$  &  $\surd$  &  $-$  &  $-$  \\
MS bifurcation/filaments   &  $(\surd)$  &  $-$  &  $-$  &  $\surd$  \\
MS density gradient          &  $-$  & $\surd$ & $(\surd)$ & $(\surd)$ \\
Metallicity of MS tip           & $(\surd)$ &  $-$  & $(\surd)$  & $(\surd)$ \\
Magellanic Bridge             & $\surd$ &  $-$  & $(\surd)$  & $\surd$ \\
Leading material exists    &  $\surd$  &  $-$  &  $\surd$  &  $\surd$  \\
LA location/kinematics     &  $-$  &  $-$  &  $-$  &  $-$  \\
SMC stellar kinematics     &  $-$  &  $-$  &  $-$  &  $\surd$  \\
\enddata

\tablenotetext{0}{The symbol ``$-$" means that the model does not explain or does not address the given property; ``$\surd$'' means that the model explains the property; and ``$(\surd)$'' means that the model can explain the property only somewhat.  Choices in each category are clarified in the text, Sec~\ref{sec:dx-compare}.}
\tablenotetext{1}{``C06" is representative of the traditional tidal models: Connors et al. (2006); Yoshizawa \& Noguchi (2003); Gardiner \& Noguchi (1996).  These models are very similar, as they adopt the same orbits for the MCs and use the same isothermal halo for the MW.  ``M05" stands for Mastropietro et al. (2005).  ``B10" stands for Besla et al. (2010) { as well as Besla et al. (2012). } }

\end{deluxetable}

The B10 model is not \emph{fully} self-consistent, however, because the Milky Way is modeled by a fixed potential.  M05 self-consistently model the interaction between the LMC and Milky Way without resorting to external potentials, but we emphasize that the SMC does not exist in their simulation.  The M05 model is the only one to include hydrodynamical interactions with the hot halo of the Milky Way, and the incorporation of this interaction into current tidal models is indeed a salient need (Sec~\ref{sec:dx-la}).  The ultimate goal to which all these models strive is the construction of a fully self-consistent model of the gravitational \emph{and} hydrodynamical interaction between the LMC, SMC, and Milky Way that (i) reproduces all MC orbital constraints, (ii) models the Milky Way realistically, and (iii) explains the MS, Bridge, and LA.  Even constructing a model that satisfies (i) and (ii) has yet to be achieved, which means that we must use far simpler models (e.g., the current work) to investigate the formation of the MS.  { To be fair, the current model is not a satisfying representation of the Magellanic system, owing to various simplifying assumptions including the neglect of dynamical friction between the MCs and failing to represent the LMC as a self-consistent N-body system.}

C06 model the Milky Way as an isothermal sphere whereas the other three models in the current comparison choose NFW halos for the Milky Way.  Owing to the prevalence of CDM cosmology, the C06 choice cannot be considered realistic.  Unlike the present work in which a bulge and disk are also ascribed to the Milky Way, the potential adopted by B10 is a lone NFW halo with a virial mass of  $1.5 \times 10^{12}$ M$_{\odot}$ and a virial radius of 240 kpc.  These parameters fall in between those of the presently adopted NFW halo, $M_{\rm vir} = 1.30 \times 10^{12}$ M$_{\odot}$, $R_{\rm vir} = 175$ kpc, and the halo of the ``alternate" model, $M_{\rm vir} = 1.90 \times 10^{12}$ M$_{\odot}$, $R_{\rm vir} = 269$ kpc (see Table~\ref{tab:param1}).  These choices for the Milky Way should be compared with mass estimates of the Milky Way at large radii.  For example, Gnedin et al. (2010) estimate that the total mass of the Milky Way within 80 kpc is $6.9^{+3.0}_{-1.2} \times 10^{11}$ M$_{\odot}$.  To compare, the total Milky Way mass within 80 kpc in our adopted model is $8.7 \times 10^{11}$ M$_{\odot}$ and in our alternate model is $9.1 \times 10^{11}$ M$_{\odot}$.  Both of these figures fall within the 1$\sigma$ uncertainty of the measured quantity.  In this sense, the adopted potential of the Milky Way in the present work is realistic, and similar arguments hold for B10 and M05.

The observed bifurcation of the MS into two distinct, parallel filaments is strongly reproduced in the present work, and we have put much effort into revealing the origin of this feature (Sec~\ref{sec:disk-bif}).  C06 previously claimed to reproduce the MS bifurcation, but it must be noted that the C06 on-sky MS distribution does not convincingly reproduce the location or morphology of the MS filaments.  Neither do they identify a plausible dynamical origin for the bifurcation.  Nevertheless, we can conservatively state that the C06 model exhibits filamentary substructure within the MS, and moreover, that the M05 and B10 models not not exhibit any such substructure.  The MS in the B10 model exhibits a brief on-sky split, but the location and kinematics of this structure do not correspond well with observations.  For this reason, we speculate that a filamentary MS may be attributable to tidal interaction over multiple passages of the Milky Way, as this is a shared property of the present work and C06 but neither M05 or B10.

{ Alternatively, Nidever et al. (2008) suggest that the filaments of the MS may have arisen from an internal ``blowout" mechanism that expelled gas from the MCs.  Using an intricate analysis of HI radial velocity profiles, they trace one of the filaments to an origin within the SMC, and they trace the other filament (and portions of the LA) to an origin within the LMC.  The present model conflicts with these results, particularly because we are able to explain the dual filaments by invoking a tidal origin within the SMC only.  Nevertheless, one of the MS filaments ``appears" to pass through the LMC in our model as judged from the right hand panel of Figure~\ref{fig:disk-rain}, and we show in Sec~\ref{sec:disk-bif} that this filament is connected to a branch of the LA which also ``appears" to overlap with the LMC in Fig~\ref{fig:disk-rain}.  Though superficially consistent with Nidever et al. (2008) in the sense that the MS filament and LA seem to be connected to the LMC, this coincidence is merely a projection effect in our model.  Regardless, we will consider possible hydrodynamical interactions between the MS/LA and the gas disk of the LMC in future work.

A possible observational test of the ``blowout" hypothesis of Nidever et al. (2008) is the measurement of metallicity along each of the MS filaments and the LA.  Because tidal models such as the present work assume the MS and LA originate from the SMC, the metallicities of these structures should be correspondingly low, as verified at the MS tip by Fox et al. (2010).  In contrast, the blowout model requires much of the stripped gas to be enriched, as would be expected from gas originating in the LMC.  In particular, the two filaments of the MS should have markedly different metallicities if one originated in the LMC and the other in the SMC as suggested by Nidever et al. (2008).  Such metallicity measurements have not been conducted but would provide valuable insight into the origin of the HI structures of the Magellanic system.  Another property of the MS filaments that bears discussion is that their kinematics exhibit sinusoidal oscillations in the analysis of Nidever et al. (2008).  This is not reproduced in the current model (Fig~\ref{fig:disk-rhovlsr}), and attempting to resolve this discrepancy will be a topic of future work incorporating hydrodynamical interactions.}

The density of the MS is observed to steadily decline across its length until it reaches low column densities at its tip (e.g., Putman et al. 2003a; Nidever et al. 2010).  The traditional tidal models such as C06 are inconsistent with this gradient because they predict an \emph{increasing} trend with a dense plume at the MS tip.  Such a dense region exists in the present work as well, but it is much more localized and located further along the MS beyond its classical tip { (see Appendix)}.  Moreover, the present model does indeed predict a generally decreasing trend in density until the tip is reached.  Similarly, the B10 model predicts the correct column densities for the MS though not the correct gradient.  Only the M05 model strongly reproduces the MS density gradient, implying that hydrodynamical interactions are needed to explain this property.  DB11b showed that the dense plume at the MS tip in the traditional tidal scenario can effectively be removed if ram pressure from the Milky Way's hot halo is taken into account.  Accordingly, it seems that the density of the MS cannot be explained by tidal models without external gas dynamical interactions with the Milky Way.

The recent observational results of Fox et al. (2010) indicate that the chemical abundances of the MS tip are rather low (e.g., [O/H]$\sim -1.00$ and [N/H]$<-0.44$), much lower than the metal-rich gas of the LMC and even lower than the chemical abundances of the SMC (e.g., [O/H] $\sim -0.66$).  Considering this measurement, Fox et al. (2010) suggest that the MS likely originated in the SMC rather than the LMC.  This is a potentially serious problem for M05 because they assume an LMC origin for the MS, in contrast to the tidal models (C06, B10, and the present work) which assume an SMC origin.  In these tidal models, the outer parts of the SMC are preferentially stripped to become the MS (e.g., Fig~\ref{fig:origin}).  If the SMC has a negative radial gradient in metallicity (i.e., if the outer gas disk is metal-poor in comparison to the central body of the SMC), then each of the tidal models would be consistent with the chemical abundance measurements of Fox et al. (2010).

The Magellanic Bridge is well reproduced in C06 and in the present work, but the gaseous structure which connects the MCs in the B10 model is displaced from the observed location of the Bridge.  Because M05 exclude the SMC from their model, they cannot address the formation of the Bridge.  The M05 model also fails to predict any stripped material on the leading side of the orbit, in contrast to the leading structures predicted by the tidal models.  Indeed, the observed presence of the LA has long been considered to support a tidal origin for the MS rather than a ram pressure origin (Putman et al. 1998).  However, the observed location, kinematics, and morphology of the LA have not been successfully explained by \emph{any} dynamical model.  This would seem to suggest that the tidal models need to incorporate hydrodynamical effects in order to account for the properties of the LA, as discussed in Sec~\ref{sec:dx-la}.

The predicted properties of the leading material may be able to distinguish between the different tidal models, even if the correspondence to observation is poor.  The C06 model does the best job in capturing the on-sky ``kink" in the LA as it extends away from the MCs, but the observed LA location is more than 30$^{\circ}$ away from the C06 prediction.  This "kink" is not exhibited in the present work, though the tip of the predicted leading material does coincide with a portion of the observed LA (Fig~\ref{fig:disk-coldens}).  The situation is worse for the B10 model, as it predicts the LA to emanate from the far side of the LMC where no HI structures are observed.  In addition, the B10 model is unable to predict an elongated leading structure possibly because the potential of the Milky Way is unable to play a vital role within the short timescale ($<0.2$ Gyr) of a first passage orbit.  The current model predicts two tidal branches within the leading material, in contrast to the single leading arms of B10 and C06.  These multiple branches may correspond to the proliferation of cloudlets observed in the LA (e.g., Diaz \& Bekki 2011a), but more extensive modeling is needed to seriously consider this point.  

One of the principal disadvantages of the C06 and B10 models is that they cannot reproduce the observed stellar kinematics of the SMC.  Neither can the M05 model, which excludes the SMC entirely.  C06 and B10 represent the SMC as a rotating disk galaxy whereas the observations of HZ06 clearly show that the red giants of the SMC are distributed in a dispersion-supported spheroid.  In the present work, we have self-consistently reproduced the kinematical differences between the gaseous and stellar components of the SMC by adopting a multi-component system composed of a disk (for the gas) and a spheroid (for the old-type stars).  As shown previously, the tidal evolution of the spheroid can also explain the observed stellar structure in the periphery of the SMC and therefore be more consistent with observations.  This suggests that future tidal models will need to adopt disk-plus-spheroid systems for the SMC rather than the traditional choice of a pure disk.

\subsection{The Magellanic past before the formation of the MS} \label{sec:dx-past}

In the present study we have proposed that the MCs have become binary companions only recently, and that their first strong tidal interaction $\sim2$ Gyr ago was signaled by the formation of the MS.  There are two general scenarios that can accommodate the MCs commencing their tidal interaction only recently.  First, the MCs may have orbited as independent satellites of the Milky Way at early times but have only recently joined as a binary pair.  This scenario is invoked by a number of studies to explain the formation of the MS (DB11a and the present work) and to solve the ``Age Gap" problem for globular clusters of the LMC (Bekki et al. 2004).  The second possibility is that the MCs were accreted onto the Milky Way as a binary pair, but the mutual separation between the MCs has become small only recently owing to the gradual decay of orbits from dynamical friction.  One realization of such an orbit has been investigated by Besla et al. (2010; see also Besla et al. 2012).  It is currently unclear which of the above two possibilities is more plausible for the origin of the MCs, due in part to the lack of \emph{fully} self-consistent simulations (i.e., representing both the MCs and Milky Way with N-body particles) that successfully reproduce the MC orbital constraints (e.g., see Table~\ref{tab:compare}).

Another important aspect of the adopted orbital model is that the MCs interact with the Milky Way during two pericenters, $\sim$2.4 Gyr ago and the present day.  We speculate that the MCs must have executed at least two passages about the Milky Way in order to reproduce the HI observations of the Magellanic system, particularly the MS bifurcation and the elongated LA (in contrast, consider the first passage models of Besla et al. 2010 and Besla et al. 2012).  Recent cosmological simulations of the LMC orbital history indicate that a first passage orbit is not necessarily the preferred scenario as long as the virial mass of the Milky Way is $\gtrsim 1.8 \times 10^{12}$ M$_{\odot}$ (Sales et al. 2011).  In the present work, the ``alternate" model for the Milky Way satisfies this requirement, as it has a virial mass of $1.9 \times 10^{12}$ M$_{\odot}$ (see Table~\ref{tab:param1}).  The total mass of the ``adopted" Milky Way model within $r<300$ kpc is comparably large, totaling $1.7 \times 10^{12}$ M$_{\odot}$.  Accordingly, we interpret our orbital model (and the preference for two pericentric passages about the Milky Way) as being broadly consistent with the cosmological simulations of Sales et al. (2011).

{ To consider the orbital history of the MCs at early times, one should not use fixed potentials as in the present work because the orbit integrations are realistic only for the past $\sim4$ Gyr or so.  Recently, a number of studies have explored more reliable methods for addressing the full orbital history of the LMC: Bekki (2011b), who uses a self-consistent treatment to investigate the accretion of the LMC onto the Milky Way from outside its virial radius, as well as Sales et al. (2011) and Boylan-Kolchin et al. (2011, hereafter BK11), who identify ``LMC-type" galaxies within large-scale cosmological simulations (see also Lux et al. 2010).  Bekki (2011b) determine that the orbital models adopted in the traditional formation scenarios for the MS (e.g., GN96) are not inconsistent with the conclusions of BK11, and we arrive to a similar conclusion for the present model.  The LMC has a number of orbital properties in the present work including (i) it is strongly bound to the Milky Way, (ii) the orbital eccentricity is $e\sim 0.5$, and (iii) the LMC has been orbiting the Milky Way for at least 4 Gyr.  Property (iii) implies that the LMC needs to be accreted onto the Milky Way more than $\sim$4 Gyr ago, and BK11 determine that such galaxies (i.e., LMC-analogs that cross the Galactic virial radius at $t_{\rm fc}>4$ Gyr ago) may have small eccentricities and are more likely to be strongly bound.  In particular, BK11 show that about 50\% of such galaxies also have eccentricities $e$ smaller than 0.6.  This result is consistent with the above properties (i) and (ii) of the present work.  We therefore conclude that the present orbital model can be consistent with the predictions of cosmological simulations as reported by BK11.}

\section{Conclusions} \label{sec:con}

We have constructed new N-body models for the tidal evolution of the SMC in a $\sim3$ Gyr interaction history with the LMC and Milky Way, and we have thereby investigated the formation processes of the MS, LA, and Bridge.  The framework of the present investigation has a number of advantages over previous tidal models of the SMC (e.g., C06, B10), including (i) our adopted orbital model was selected after conducting a large parameter space search centered on the measured proper motion values of Vieira et al. (2010), (ii) the Milky Way is represented by a realistic potential having a bulge, disk, and NFW halo, and (iii) we investigated the stellar kinematics and outer stellar structure of the SMC by adopting ``disk plus spheroid" models in good agreement with observations (e.g., Harris \& Zartisky 2006; Nidever et al. 2011).  Our main results are summarized as follows.

(1) The LMC and SMC become dynamically coupled only recently, suffering two strong tidal interactions $\sim2$ Gyr ago and $\sim250$ Myr ago.  Portions of the SMC disk are stripped away during each of these strong encounters, with the MS and LA being created during the first encounter, and the Bridge during the second.  The timing of these two tidal interactions is consistent with the correlated star formation epochs determined by Harris \& Zaritsky (2009).

(2) The morphology of the MS is well reproduced, including its bifurcation into dual filaments (Putman et al. 2003a) and its long extent (Nidever et al. 2010).  The predicted on-sky location of the filaments provides a very strong match to observations, as does their crossing point.  The kinematics of the MS are also well reproduced, particularly the velocity inflection at the MS tip (Nidever et al. 2010).

(3) We interpret the observed MS bifurcation as a wrapping of the original tidal arm (``tidal arm A") that was stripped from the SMC disk.  The dynamical process responsible for this ``wrapping" can be traced to a complex three-body interplay between the SMC, LMC, and MW over the past $\sim2$ Gyr.  We accordingly suggest that the HI morphology of the MS places strong constraints on the orbital history of the MCs, namely that the MCs have executed at least two pericentric passages about the MW during a $\sim2$ Gyr bound association.

(4) The formation of the Bridge $\sim250$ Myr ago was accompanied by the formation of a complementary structure called the ``Counter-Bridge" that extends $\sim20$ kpc along the line of sight behind the SMC.  Because of this alignment, the Counter-Bridge does not appear as an identifiable structure in column density maps, but it is likely traced by the large line-of-sight depth of the SMC (e.g., Crowl et al. 2001; Groenewegen 2000).

(5) Two branches of leading material are stripped from the SMC disk, but the on-sky location and kinematics do not correspond well with the observed properties of the LA. Even though we do not explicitly account for
hydrodynamical effects in our simulation, we have argued that the evolution of the leading material over the past $\sim300$ Myr should be strongly influenced by ram pressure from the hot halo of the Milky Way (Section~\ref{sec:dx-la} and Fig~\ref{fig:rvt}).  Accordingly, we suggest that gravity alone cannot explain the present location and kinematics of the LA, even if this structure is tidal in origin.

(6) The LMC engulfs a large fraction of the material that is tidally stripped from the SMC following their strong tidal interactions $\sim2$ Gyr ago and $\sim250$ Myr ago.  The transfer of SMC disk material to the LMC would correspond to the infall of metal-poor gas and would be imprinted into the chemical enrichment history of the LMC (e.g., { Fig 1 of van Loon et al. 2005}; Fig 20 of Harris \& Zaritsky 2009; Rubele et al. 2012) as well as the nitrogen-poor stellar populations of the LMC (e.g., Bekki \& Tsujimoto 2010).  The mass transfer from the SMC spheroid would correspond to the accretion of old-type stars onto the LMC and is suggestive of various observed populations of kinematically peculiar stars (e.g., Kunkel et al. 1997a, Olsen et al. 2011).

(7) Of our three models for the SMC spheroid, we find that the extended spheroid is best able to reproduce the relevant observations, including the stellar kinematics of the SMC (Harris \& Zaritsky 2006), the recent discovery of a break population of red giants (Nidever et al. 2011), and the observation of kinematically and chemically peculiar stars within the LMC (Olsen et al. 2011).  Considering these results, future observations may need to readdress the possibility that SMC stars were tidally stripped into the Bridge region.

(8) The tidal evolution of the extended spheroid predicts that a stellar stream should be stripped away $\sim2$ Gyr ago { at the same time that the MS is stripped from the SMC disk.  Whereas the gaseous MS originates in the rotating SMC disk, the stellar stream originates in the pressure-supported spheroidal component of the SMC.  This distinct origin suggests that the stripped stars should not correlate with the peak HI regions of the MS, as observed (e.g., Recillas-Cruz 1982; Bruck \& Hawkins 1983; Guhathakurta \& Reitzel 1998).  Future wide-field surveys such as the Southern Sky Survey by the SkyMapper telescope (Keller et al. 2007) would be ideal for detecting this possible tidal stream and establishing its connection to the stellar halo of the SMC.}

\acknowledgements

J.D.D. acknowledges the financial support of the Gates Cambridge Trust during the completion of this paper and a stipend from ICRAR at the University of Western Australia during the initial phase of the research.  K.B. acknowledges the financial support of the Australian Research Council throughout the course of this work.  We would like to thank David Nidever and Mary Putman for stimulating conversations and for allowing us to incorporate their data into the paper; Katherine Vieira and Gurtina Besla for clarifying discussions on their recent work; Tobias Westmeier, Simone Zaggia, Jacco van Loon, Tim Connors, and Christian Bruns for their comments and insight; Lee Kelvin and Stefan Westerlund for software support; and Paul Bourke of the iVEC facility at the University of Western Australia for lending his expertise in real-time visualization.  Lastly, we thank the anonymous referee for constructive comments that led to the improvement of the manuscript. Numerical computations reported here were carried out on the GRAPE system at ICRAR in the University of Western Australia.\\



\appendix

In this paper we have presented an N-body model for the Magellanic system characterized by the set of parameter values given in Tables 1 and 2.  Here we briefly discuss parameter choices other than those adopted to give a sense of the parameter space that was explored.  In particular, we will discuss Galactic parameters, proper motions of the MCs, masses of the MCs, disk angles for the SMC, and the initial time of the simulation.

\begin{figure} \centering
\includegraphics[trim=0 0 0 0, scale=0.4]{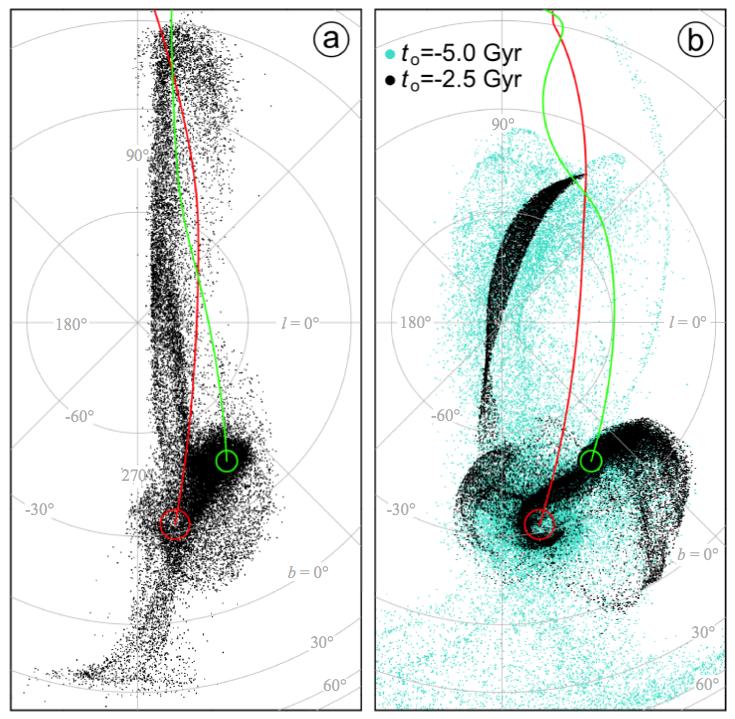}
\caption{The disruption of the SMC disk with different parameter choices. (a) On-sky projection of the ``alternate model" (see Table~\ref{tab:param1} and Sec~\ref{sec:mod-orb}) which differs from the adopted model in the choice of parameters for the Milky Way halo.  The N-body model is a pure disk (Sec~\ref{sec:mod-nbody-pure}) having $N=10^5$ total particles and disk angles $\theta_{\rm d} = -40^{\circ}$ and $\phi_{\rm d} = 230^{\circ}$. (b)  On-sky projection of test particle models ($N=10^5$; see Sec~\ref{sec:mod-test}) with proper motions taken from Diaz \& Bekki (2011a), which is consistent with the measurements of Kallivayalil et al. (2006a,b).  All other parameters are taken from Table~\ref{tab:param1}, except for the mass of the LMC, which is taken to be $2\times10^{10}$ M$_{\odot}$ as in Diaz \& Bekki (2011a).  The test particle models differ in the time at which each simulation is begun, $t=-2.5$ Gyr (black) or $t=-5.0$ Gyr (turquoise).  Coordinates in both panels are galactic longitude $l$ (straight lines) and latitude $b$ (concentric circles) shown in a ZEA projection centered on the South Galactic Pole.  The orbit trails of the SMC (green) and LMC (red) are also shown, as are their current locations (circles).
\label{fig:app-lb}}
 \end{figure}

Panel (a) of Figure~\ref{fig:app-lb} gives the on-sky projection of the ``alternate model" as described in Sec~\ref{sec:mod-orb}, realized as a pure-disk N-body model (Sec~\ref{sec:mod-nbody-pure}).  The parameter values of this N-body model are the same as those of the adopted model except for a different characterization of the Milky Way's NFW halo (see Fig~\ref{fig:rot} and footnote of Table~\ref{tab:param1}).  These two models share a number of properties, including the fact that their tidally stripped material has nearly identical morphology and kinematics.  This similarity in morphology can be assessed by comparing panel (a) of Fig~\ref{fig:app-lb} with the left-hand panel of Fig~\ref{fig:disk-coldens}.  Of particular importance is that the MS exhibits a bifurcation into two distinct filaments in both models.  The overall kinematics of the MS including the velocity inflection (Nidever et al. 2010) are also reproduced in the alternate model, though we do not show the kinematics here. 

The alternate model highlights a strength of the present study, namely that the formation of the MS is consistent with multiple prescriptions for the Milky Way halo.  The model is therefore more robust in the face of observational uncertainties for a number of Galactic parameters, including the mass of the Milky Way (e.g., Gnedin et al. 2010) and the circular velocity $V_{\rm cir}$ at the solar radius (e.g., Kerr \& Lynden-Bell 1986; Reid et al. 2009).  For instance, we find that values of $V_{\rm cir}$=240 km/s (adopted model) and $V_{\rm cir}$=220 km/s (alternate model) are consistent with the formation of the MS without prejudice or preference for either value.  Though we found only two possible models in our work, it is likely that there exist many other parameterizations of the Milky Way's halo that can generate similar models for the MS.  Furthermore, it would be worthwhile to consider whether other halo profiles (i.e., other than NFW) can generate similar models.

\begin{figure} \centering
\includegraphics[trim=55 50 0 370, scale=0.5]{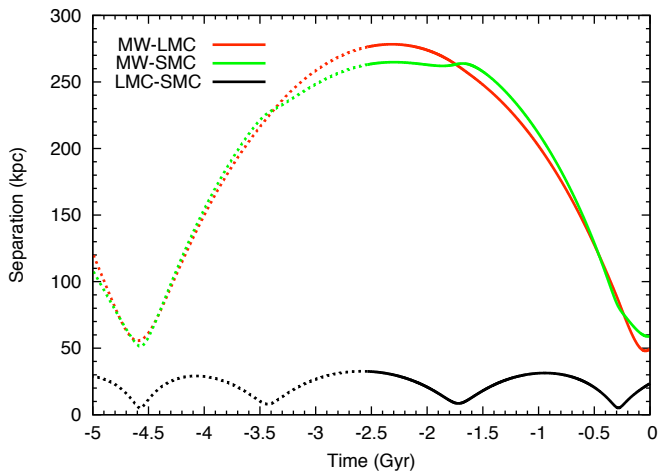}
\caption{Orbital separations between the LMC and MW (red), SMC and MW (green), and LMC and SMC (black) for the two test particle models given in panel (b) of Figure~\ref{fig:app-lb}.  Solid lines indicate the time window from $t=-2.5$ Gyr to the present day, whereas the dotted lines begin at an initial time of $t=-5.0$ Gyr.
\label{fig:app-sep}}
 \end{figure}

The present model is less flexible, however, when considering different proper motion values for the LMC and SMC, especially those of Kallivayalil et al. (2006a,b; K06).  Panel (b) of Figure~\ref{fig:app-lb} provides the on-sky distribution of two test particle models (Sec~\ref{sec:mod-test}) using the proper motions of Diaz \& Bekki (2011a), which is within 1$\sigma$ of the K06 measured values (see Figure~\ref{fig:pm}).  The black (turquoise) particles are evolved to the present day from an undisturbed SMC disk at $t=-2.5$ Gyr ($t=-5.0$ Gyr), and the orbital separations are given as the solid (dashed and solid) lines in Figure~\ref{fig:app-sep}.  As is clear, both cases (black and turquoise) give poor results.  The stripped material does not provide a morphological match to any of the HI structures of the Magellanic system, and the kinematics (not shown) are similarly unacceptable.  The inability of our adopted model to accommodate the proper motions of K06 is perhaps unsurprising, since the orbital interaction history changes dramatically (e.g., compare Fig~\ref{fig:sep} and Fig~\ref{fig:app-sep}) owing to the substitution of much larger proper motions.

\begin{figure} \centering
\includegraphics[scale=0.55, trim=0 100 0 70]{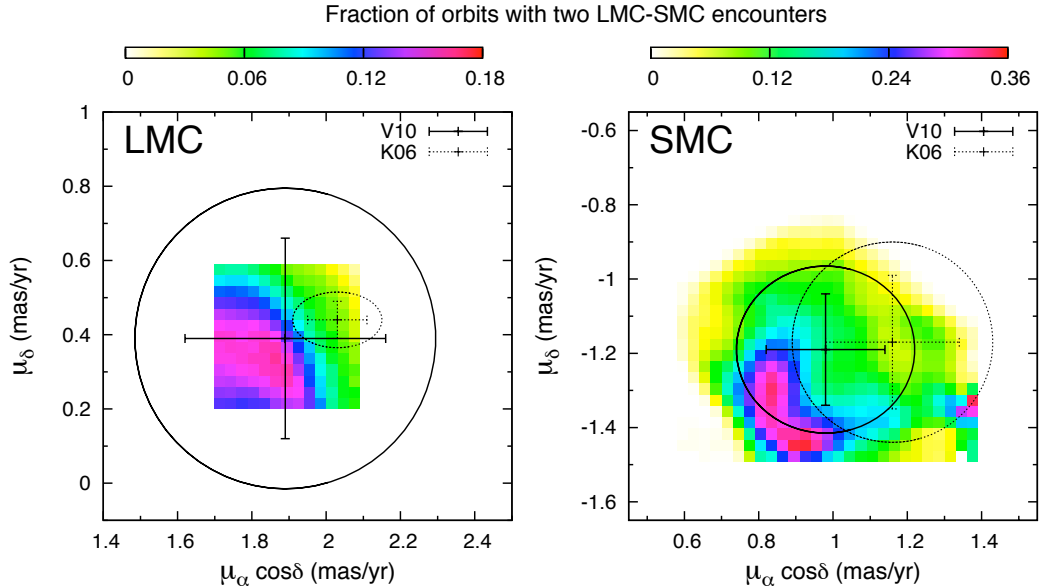}
\caption{The fraction of orbits having exactly two encounters between the LMC and SMC, plotted as a function of the proper motions of the LMC (left) and SMC (right).  In each panel, we consider all $\sim 8 \times 10^6$ orbits computed in the present study, and the fraction is relative to the total number of orbits computed at each proper motion value.  For example, at each point in the LMC proper motion plane, there are $\sim5 \times 10^4$ total orbital models, and the color indicates the fraction of these which satisfy the condition of two encounters.  An ``encounter" is defined as an approach of less than 10 kpc that is also a minimum in the LMC-SMC separation curve (e.g., Fig~\ref{fig:sep}, Fig~\ref{fig:app-sep}).  See discussion in the Appendix indicating that the condition of two LMC-SMC encounters is a promising criterion for the formation of the MS.  Observational ellipses outline the 68.3\% confidence regions and 1$\sigma$ error bars for Vieira et al. (2010; V10, solid) and Kallivayalil et al. (2006a,b; K06, dashed).
\label{fig:app-pms}}
 \end{figure}

Our decision to establish our parameter space around the measured proper motion values of Vieira et al. (2010, V10) rather than those of K06 was informed by a number of factors, including the failure of K06 test particle models similar to those shown in panel (b) of Fig~\ref{fig:app-lb}.  Moreover, as mentioned in Sec~\ref{sec:int}, the V10 study possess a number of advantages over K06, including a larger sample size, longer baseline, and the ability to measure \emph{relative} proper motions between the LMC and SMC.  Lastly, our preference for V10 was also led by the overall properties of the orbital models.  For instance, Figure~\ref{fig:app-pms} gives the fraction of orbits (within the full set of $8\times10^6$ explored) in which the SMC suffers exactly two close encounters with the LMC.  As emphasized below (see also Sec~\ref{sec:mod-orb}), there is mounting evidence that such a condition on the orbital history is preferred for the formation of the MS (e.g., Besla et al. 2010; Ruzicka et al. 2010; Diaz \& Bekki 2011a; present work).  Of course, there are many more constraints that must be considered (e.g., see Sec~\ref{sec:mod-orb} and Sec~\ref{sec:mod-test}), but the condition of exactly two LMC-SMC encounters is a good starting point for considering which regions of parameter space are promising for MS formation scenarios.  Fig~\ref{fig:app-pms} clearly shows that the proper motions with the largest fraction of promising orbits are much closer to the V10 values than those of K06.  This implies that a parameter space centered on V10 may be more likely to harbor a successful formation scenario for the MS, as in the present work.  This conclusion is bolstered by the work of Ruzicka et al. (2010), who could not find a compelling model for the MS within a parameter space centered on the K06 proper motions.  It is also striking to note that the proper motions of our adopted model (Fig~\ref{fig:pm}) are close to the peak regions in Figure~\ref{fig:app-pms}.

Despite the conclusions of the present work, it may still be possible to reconcile the K06 values with a viable MS formation scenario.  Furthermore, even though we could not find a first passage model that could explain the observations of the MS, the first passage scenario is not necessarily ruled out.  It would seem that a different set of assumptions (i.e., differing from those of Sec~\ref{sec:mod}) would be needed to build a model with the K06 proper motions.  As examples, consider the massive isothermal halo for the Milky Way adopted by Diaz \& Bekki (2011a) or efficient dynamical friction between the MCs utilized by Besla et al. (2010; see also Besla et al. 2012).   One would also need to investigate different masses for the LMC and SMC.  For instance, Besla et al. (2010) assume that the LMC and SMC have halo masses of 1.8$\times 10^{11}$ M$_{\odot}$ and 2.5$\times 10^{10}$ M$_{\odot}$, respectively, which are roughly an order of magnitude larger than adopted in the present study (Table~\ref{tab:param1}).  The reason behind this large difference in adopted MC masses may possibly be traced to the difference in model assumptions between the two studies.  In particular, mutual dynamical friction between the MCs is ignored in the present study, but in Besla et al. (2010), it is the mechanism that drives the MCs toward the mutual encounter that creates the MS.  It is not surprising that a different range of masses for the LMC and SMC are preferred in the presence/absence of dynamical friction, especially because the magnitude of the effect is sensitive to satellite mass (e.g., see equation (\ref{chandra})).

In the present work, we explored a range of masses similar to those of traditional tidal models (e.g., GN96, C06): (1.0, 1.5, 2.0, 2.5, 3.5, 4.0) $\times 10^{10}$ M$_{\odot}$ for the LMC, and (3.0, 4.5) $\times 10^9$ M$_{\odot}$ for the SMC.  We could find promising formation scenarios for the MS with only one mass combination (i.e., 1.0$\times 10^{10}$ M$_{\odot}$ and 3.0$\times 10^9$ M$_{\odot}$), which certainly falls in the ``low mass" regime for the MCs (e.g., van der Marel et al. 2002, Stanimirovi\'c et al. 2004, Bekki \& Stanimirovi\'c 2009).  This may be a reflection of the actual masses of the MCs (or possibly their mass ratio), but one must not forget that the adopted values are a product of our particular model assumptions, just as in the case of large MC masses for Besla et al. (2010).  Among our assumptions is the fact that the LMC is represented by a fixed Plummer potential (Sec~\ref{sec:mod-orb}), which underpins the lack of dynamical friction between the MCs.  For the LMC itself, this assumption is not unreasonable since its mass is unlikely to change within the past $\sim$2 Gyr, as verified by live models of the LMC (e.g., Bekki \& Chiba 2005).  It remains unclear, however, if our neglect of dynamical friction between the MCs is realistic.

\begin{figure} \centering
\includegraphics[trim=50 50 50 400, scale=0.6]{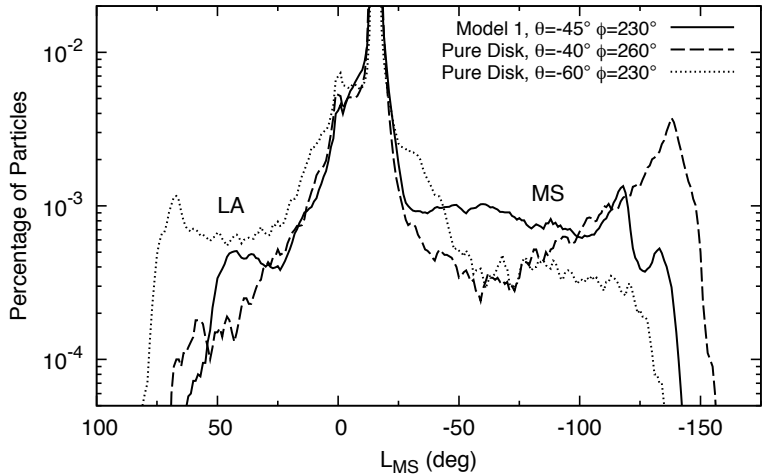}
\caption{Particle count as a function of Magellanic longitude ${\rm L}_{\rm MS}$ quoted as a percentage of the total for three models: the disk of Model 1 (solid $\theta_{\rm d}=-45^{\circ}$ $\phi_{\rm d}=230^{\circ}$) and two different pure disk models (dashed $\theta_{\rm d}=-40^{\circ}$ $\phi_{\rm d}=260^{\circ}$; and dotted $\theta_{\rm d}=-60^{\circ}$ $\phi_{\rm d}=230^{\circ}$).  See Sec~\ref{sec:mod-nbody-pure} for the definition of pure disk models.  The density profile of the MS in each model is given at ${\rm L}_{\rm MS}<-25^{\circ}$, and that of the LA at ${\rm L}_{\rm MS}>0^{\circ}$.
\label{fig:app-cnt}}
 \end{figure}

Figure~\ref{fig:app-cnt} shows how our N-body model is affected by adjusting the SMC disk angles $\theta_{\rm d}$ and $\phi_{\rm d}$ (see definitions in Sec~\ref{sec:mod-nbody-pure}).  In general, we found that the best disk angles fell in the range $-60^{\circ} < \theta_{\rm d} < -40^{\circ}$ and $220^{\circ} < \phi_{\rm d} < 260^{\circ}$, and models outside of this range failed to reproduce a tidal stream resembling the MS.  Even within this range, however, many models provide a poor match to the MS, with either too many particles at the MS tip (Fig~\ref{fig:app-cnt}, dashed line) or too few particles in the MS altogether (dotted).  The reasons for this sensitivity to disk orientation have not yet been fully explored, nor is it clear why there seems to be a preference for $\theta_{\rm d}=-45^{\circ}$ and $\phi_{\rm d}=230^{\circ}$ regardless of tidal history (i.e, these values were found to be best in GN96 as well as the present work).  Although the density profile of the MS in our adopted model (Fig~\ref{fig:app-cnt}, solid line) is superior to other candidate models, there are discrepancies with observation.  The column density of the MS is observed to decline exponentially along its length (e.g., Figure 10 of Nidever et al. 2010), whereas the particle count in the adopted model exhibits a small peak at the MS tip ${\rm L}_{\rm MS} \approx -120^{\circ}$, and furthermore the density along the MS does not decline as rapidly as observed.  Reconciling theory with observation will likely require the inclusion of hydrodynamics (Sec~\ref{sec:dx-la}).

The final input parameter that we discuss is the initial time of the simulation.  Despite the poor observational correspondence of the test particle models in panel (b) of Fig~\ref{fig:app-lb}, the model that evolves from $t=-2.5$ Gyr (black) is able to form coherent tidal structures in contrast to the model that evolves from $t=-5.0$ Gyr (turquoise) whose structures are ``smeared" spatially and kinematically.  This difference can be traced to the interaction history given in Fig~\ref{fig:app-sep}.  The SMC suffers two close encounters with the LMC in the black model (solid lines), similar to the scenario in the present work.  The turquoise model evolves through these interactions as well, but only after suffering earlier encounters with the LMC and MW between $t=-5.0$ Gyr and $t=-2.5$ Gyr (dashed lines).  The multiplicity of these interactions effectively destroys the coherence of the stripped material.  Similarly, GSF94 find that the coherence of the MS is destroyed if the SMC is subjected to tidal interactions previous to $\sim$3 Gyr ago.  Based on these results and those of other recent models (e.g. Besla et al. 2010; Ruzicka et al. 2010; Diaz \& Bekki 2011a), we suggest that the creation of a coherent tidal structure resembling the observed MS requires \emph{exactly two} recent encounters between the LMC and SMC.

The time at which the simulation is begun is therefore an important parameter for models having earlier epochs of strong tidal interaction (e.g., GSF94, GN96, C06).  For example, if the LMC and SMC are indefinitely bound to one another (as in the models of panel (b) in Fig~\ref{fig:app-lb}), one may artificially ensure that only two strong interactions occur by appropriately choosing the initial time for the simulation.  This tactic is unnecessary for models that assume the LMC and SMC have only recently entered their bound state (Besla et al. 2010; Diaz \& Bekki 2011a; present work).  A recent dynamical coupling between the LMC and SMC can elegantly deliver the two necessary tidal encounters for the formation of the MS while keeping the SMC safe from external tidal fields at earlier epochs.  In principle, the starting time of the simulation should not matter for such models as long as it is prior to the first strong encounter.  We mention briefly a caveat of the present model, however, because the initial time of our adopted model is chosen at $t=-3.37$ Gyr.  We found that the internal structure of the MS depends somewhat on the orientation of the SMC bar at the first epoch of tidal stripping, and since the phase of the SMC bar depends on the initial start time of the simulation, the present results depend weakly on the initial time.


\begin{thebibliography}{}

\bibitem[()]{} Bekki, K., Couch, W. J., Beasley, M. A., et al. 2004, ApJ, 610, L93

\bibitem[()]{} Bekki, K. 2011, ApJ, 730, L2 (2011a)

\bibitem[()]{} Bekki, K. 2011, MNRAS, 416, 2359 (2011b)


\bibitem[()]{} Bekki, K., \& Chiba, M. 2005, MNRAS, 356, 680

\bibitem[()]{} Bekki, K., \& Chiba, M. 2007, MNRAS, 381, L16

\bibitem[()]{} Bekki, K., \& Chiba, M. 2009,  PASA, 26, 48

\bibitem[()]{} Bekki, K., \& Stanimirovi\'c, S. 2009, MNRAS, 395, 342

\bibitem[()]{} Bekki, K., \& Tsujimoto, T. 2010, ApJ, 721, 1515 

\bibitem[()]{} Besla, G., Kallivayalil, N., Hernquist, L., et al. 2007, ApJ, 668, 949

\bibitem[()]{} Besla, G., Kallivayalil, N., Hernquist, L., et al. 2010, ApJ, 721, L97 (B10)

\bibitem[()]{} Besla, G., Kallivayalil, N., Hernquist, L., et al. 2012, accepted to MNRAS (arXiv:1201.1299)

\bibitem[()]{} Binney, J., \& Tremaine, S. 2008, Galactic Dynamics. Princeton Univ. Press, Princeton, NJ

\bibitem[()]{} Bland-Hawthorn, J., Sutherland, R., Agertz, O., \& Moore, B. 2007, ApJ, 670, L109

\bibitem[()]{} Boylan-Kolchin, M., Besla, G., \& Hernquist, L. 2011, MNRAS, 414, 1560 (BK11)

\bibitem[()]{} Bruck, M. T., \& Hawkins, M. R. S. 1983, A\&A, 124, 216

\bibitem[()]{} Bruns, C., Kerp, J., Staveley-Smith, L., et al. 2005, A\&A, 432, 45

\bibitem[()]{} Chun, S.-H., Kim, J.-W., Sohn, S. T., et al. 2010, AJ, 139, 606

\bibitem[()]{} Cioni, M.-R. L., van der Marel, R. P., Loup, C., \& Habing, H. J. 2000, A\&A, 359, 601

\bibitem[()]{} Connors, T. W., Kawata, D., \& Gibson, B. K. 2006, MNRAS, 371, 108 (C06)

\bibitem[()]{} Costa, E., Mendez, R. A., Pedreros, M. H., et al. 2009, AJ, 137, 4339

\bibitem[()]{} Crowl, H. H., Sarajedini, A., Piatti, A. E., et al. 2001, AJ, 122, 220


\bibitem[()]{} De Propris, R., Rich, R. M., Mallery, R. C., \& Howard, C. D. 2010, ApJ, 714, L249

\bibitem[()]{} Dehnen, W., \& Binney, J. J. 1998, MNRAS, 298, 387

\bibitem[()]{} Diaz, J., \& Bekki, K. 2011, MNRAS, 413, 2015 (DB11a, 2011a)

\bibitem[()]{} Diaz, J., \& Bekki, K. 2011, PASA, 28, 117 (DB11b, 2011b)

\bibitem[()]{} Fox, A. J., Wakker, B. P., Smoker, J. V., et al. 2010, ApJ, 718, 1046

\bibitem[()]{} Freedman, W. L., Madore, B. F., Gibson, B. K., et al. 2001, ApJ, 553, 47

\bibitem[()]{} Gardiner, L. T., \& Hawkins, M. R. S. 1991, MNRAS, 251, 174

\bibitem[()]{} Gardiner, L. T., \& Noguchi, M. 1996, MNRAS, 278, 191 (GN96)

\bibitem[()]{} Gardiner, L. T. 1999, in ASP Conf. Ser. 166, Stromlo Workshop on High-Velocity Clouds, ed. B. K. Gibson \& M. E. Putman (San Francisco: ASP), 292

\bibitem[()]{} Gardiner, L. T., Sawa, T., \& Fujimoto, M. 1994, MNRAS, 266, 567

\bibitem[()]{} Gillessen, S., Eisenhauer, F., Trippe, S., et al. 2009, ApJ, 692, 1075

\bibitem[()]{} Gnedin, O. Y., Brown, W. R., Geller, M. J., \& Kenyon, S. J. 2010, ApJ, 720, L108

\bibitem[()]{} Groenewegen, M. A. T. 2000, A\&A, 363, 901

\bibitem[()]{} Guhathakurta, P., \& Reitzel, D. B. 1998, in ASP Conf. Ser. 136, Galactic Halos: A UC Santa Cruz Workshop, ed. Zaritsky, D., (San Francisco: ASP), 22

\bibitem[()]{} Gunn, J. E. \& Gott, J. R. 1972, ApJ, 176, 1

\bibitem[()]{} Harris, J. 2007, ApJ, 658, 345

\bibitem[()]{} Harris, J., \& Zaritsky, D. 2006, AJ, 131, 2514 (HZ06)

\bibitem[()]{} Harris, J., \& Zaritsky, D. 2009, AJ, 138, 1243

\bibitem[()]{} Hatzidimitriou, D., \& Hawkins, M. R. S. 1989, MNRAS, 241, 667


\bibitem[()]{} Heller, P., \& Rohlfs, K. 1994, A\&A, 291, 743

\bibitem[()]{} Hill, V. 2004, in Carnegie Observatories Astrophysics Series, Vol. 4, Origin and Evolution of the Elements, ed. A. McWilliam \& M. Rauch (Cambridge: Cambridge Univ. Press), 203

\bibitem[()]{} Irwin, M. J., Kunkel, W. E., \& Demers, S. 1985, Nature, 318, 160

\bibitem[()]{} Jin, S., \& Lynden-Bell, D. 2008, MNRAS, 383, 1686

\bibitem[()]{} Kallivayalil, N., van der Marel, R. P., Alcock, C., et al. 2006, ApJ, 638, 772 (K06, 2006a)

\bibitem[()]{} Kallivayalil, N., van der Marel, R. P., \& Alcock, C. 2006, ApJ, 652, 1213 (K06, 2006b)

\bibitem[()]{} Keller, S. C., Schmidt, B. P., Bessell, M. S., et al. 2007, PASA, 24, 1

\bibitem[()]{} Kerr, F. J., \& Lynden-Bell, D. 1986, MNRAS, 221, 1023 

\bibitem[()]{} Kim, S., Staveley-Smith, L., Dopita, M. A., et al. 1998, ApJ, 503, 674

\bibitem[()]{} Klypin, A., Zhao, H., \& Somerville, R. S. 2002, ApJ, 573, 597

\bibitem[()]{} Korn, A. J., Keller, S. C., Kaufer, A., et al. 2002, A\&A, 385, 143

\bibitem[()]{} Kunkel, W. E., Demers, S., Irwin, M. J., \& Albert, L. 1997a, ApJ, 488, L129

\bibitem[()]{} Kunkel, W. E., Irwin, M. J., \& Demers, S. 1997b, A\&AS, 122, 463

\bibitem[()]{} Lah, P., Kiss, L. L., \& Bedding, T. R. 2005, MNRAS, 359, L42

\bibitem[()]{} Lin, D. N. C., \& Lynden-Bell, D. 1977, MNRAS, 181, 59

\bibitem[()]{} Lux, H., Read, J. I., \& Lake, G. 2010, MNRAS, 406, 2312

\bibitem[()]{} Majewski, S. R., Skrutskie, M. F., Weinberg, M. D., \& Ostheimer, J. C. 2003, ApJ, 599, 1082

\bibitem[()]{} Majewski, S. R., Nidever, D. L., Mu\~noz, R. R., et al. 2009, in van Loon J. T., Oliveira M., eds, Proc. IAU Symp. 256, The Magellanic System: Stars, Gas, and Galaxies. Cambridge Univ. Press, Cambridge, p. 51

\bibitem[()]{} Mastropietro, C., Moore, B., Mayer, L., Wadsley, J., \& Stadel, J. 2005, MNRAS, 363, 509 (M05)


\bibitem[()]{} Mathewson, D. S., Ford, V. L., \& Visvanathan, N. 1988, ApJ, 333, 617

\bibitem[()]{} McClure-Griffiths, N. M., Staveley-Smith, L., Lockman, F. J., et al. 2008, ApJ, 673, L143

\bibitem[()]{} Meurer, G. R., Bicknell, G. V., \& Gingold, R. A. 1985, PASA, 7, 19

\bibitem[()]{} Miyamoto, M., \& Nagai, R. 1975, PASJ, 27, 533

\bibitem[()]{} Moore, B., \& Davis, M. 1994, MNRAS, 270, 209

\bibitem[()]{} Muller, E., \& Bekki, K. 2007, MNRAS, 381, L11

\bibitem[()]{} Mu\~noz, R. R., Majewski, S. R., Zaggia, S., et al. 2006, ApJ, 649, 201

\bibitem[()]{} Mu\~noz, R. R., Majewski, S. R., \& Johnston, K. V. 2008, ApJ, 679, 346

\bibitem[()]{} Murai, T., \& Fujimoto, M. 1980, PASJ, 32, 581 (MF80)

\bibitem[()]{} Navarro, J. F., Frenk, C. S., \& White, S. D. M. 1996, ApJ, 490, 493

\bibitem[()]{} Nidever, D. L., Majewski, S. R., \& Burton, W. B. 2008, ApJ, 679, 432

\bibitem[()]{} Nidever, D. L., Majewski, S. R., Burton, W. B., \& Nigra, L. 2010, ApJ, 723, 1618

\bibitem[()]{} Nidever, D. L., Majewski, S. R., Mu\~noz, R. R., et al. 2011, ApJ, 733, L10 (N11)

\bibitem[()]{} Noel, N. E. D., \& Gallart, C. 2007, ApJ, 665, L23

\bibitem[()]{} Olano, C. A. 2004, A\&A, 423, 895

\bibitem[()]{} Olsen, K., Zaritsky, D., Blum, R. D., Boyer, M. L., \& Gordon, K. D. 2011, ApJ, 737 29



\bibitem[()]{} Piatek, S., Pryor, C., \& Olszewski, E. W. 2008, ApJ, 135, 1024

\bibitem[()]{} Piatti, A. E., Sarajedini, A., Geisler, D., Seguel, J., \& Clark, D. 2005, MNRAS, 358, 1215

\bibitem[()]{} Piatti, A. E. 2011, MNRAS, 418, L40

\bibitem[()]{} Putman, M. E., Gibson, B. K., Staveley-Smith, L., et al. 1998, Nature, 394, 752

\bibitem[()]{} Putman, M. E., Staveley-Smith, L., Freeman, K. C., Gibson, B. K., \& Barnes, D. G. 2003, ApJ, 586, 170 (2003a)

\bibitem[()]{} Putman, M. E., Staveley-Smith, L., Freeman, K. C., et al. 2003, ApJ, 597, 948 (2003b)

\bibitem[()]{} Recillas-Cruz, E. 1982, MNRAS, 201, 473

\bibitem[()]{} Reid, M. J., \& Brunthaler, A. 2004, ApJ, 616, 872

\bibitem[()]{} Reid, M. J., Menten, K. M., Zheng, X. W., et al. 2009, ApJ, 700, 137

\bibitem[()]{} Rubele, S., Kerber, L., Girardi, L., et al. 2012, A\&A, 537, 106

\bibitem[()]{} Ruzicka, A., Theis, C., \& Palous, J. 2010, ApJ, 725, 369



\bibitem[()]{} Saha, A., Olszewski, E. W., Brondel, B., et al. 2010, AJ, 140, 1719

\bibitem[()]{} Sales, L. V., Navarro, J. F., Cooper, A. P., et al. 2011, MNRAS, 418, 648

\bibitem[()]{} Sirko, E., Goodman, J., Knapp, G. R., et al. 2004, AJ, 127, 914

\bibitem[()]{} Stanimirovi\'c, S., Staveley-Smith, L., \& Jones, P. A. 2004, ApJ, 604, 176

\bibitem[()]{} Stanimirovi\'c, S., Hoffman, S., Heiles, C., et al. 2008, ApJ, 680, 276

\bibitem[()]{} Staveley-Smith, L., Kim, S., Calabretta, M. R., Haynes, R. F., \& Kesteven, M. J. 2003, MNRAS, 339, 87

\bibitem[()]{} Swaters, R. A., van Albada, T. S., van der Hulst, J. M., \& Sancisi, R. 2002, A\&A, 390, 829

\bibitem[()]{} Subramanian, S., \& Subramaniam, A. 2009, A\&A, 503, L9

\bibitem[()]{} Subramanian, S., \& Subramaniam, A. 2012, ApJ, 744, 128

\bibitem[()]{} Toomre, A., \& Toomre, J. 1972, ApJ, 178, 623


\bibitem[()]{} van der Marel, R. P., Alves, D. R., Hardy, E. \& Suntzeff, N. B. 2002, AJ, 124, 2639

\bibitem[()]{} van Loon, J. Th., Marshall, J. R., \& Zijlstra, A. A. 2005, A\&A, 442, 597

\bibitem[()]{} van Loon, J. Th., Oliveira, J. M., Gordon, K. D., et al. 2010, AJ, 139, 68

\bibitem[()]{} Vieira, K., Girard, T. M., van Altena, W. F., et al. 2010, AJ, 140, 1934 (V10)


\bibitem[()]{} Wakker, B. P. 2001, ApJS, 136, 463

\bibitem[()]{} Welch, D. L., McLaren, R. A., Madore, B. F., \& McAlary, C. W. 1987, ApJ, 321, 162

\bibitem[()]{} Westmeier, T., \& Koribalski, B. S. 2008, MNRAS, 388, L29

\bibitem[()]{} Westmeier, T., Braun, R., \& Koribalski, B. S. 2011, MNRAS, 410, 2217


\bibitem[()]{} Yoshizawa, A. M., \& Noguchi, M. 2003, MNRAS, 339, 1135

\bibitem[()]{} Zentner, A. R., \& Bullock, J. S. 2003, ApJ, 598, 49

\end{thebibliography}
\end{document}